\documentclass[12pt]{iopart}
\usepackage{subfig}
\usepackage{graphicx}
\usepackage{epsfig}
\usepackage{epstopdf}
\usepackage[left=23mm,right=13mm,top=35mm,columnsep=15pt]{geometry} 
\usepackage{placeins}
\usepackage[T1]{fontenc}

\usepackage[sort&compress,square,numbers]{natbib}
\usepackage{iopams}
\newcommand\newblock{\hskip .11em\@plus.33em\@minus.07em}

\expandafter\let\csname equation*\endcsname\relax
\expandafter\let\csname endequation*\endcsname\relax

\usepackage{amsthm}
\usepackage{amsfonts}
\usepackage{amssymb}
\usepackage{amsmath}
\usepackage{amscd}

\usepackage[english]{babel}
\usepackage[utf8]{inputenc}
\usepackage[labelformat=parens,labelsep=quad,skip=3pt]{caption}

\usepackage{setspace}

\usepackage{bm}

\usepackage[pdftex, pdftitle={Article}, pdfauthor={Author}]{hyperref} 

\begin{document}
\title{Smith-Purcell radiation of a vortex electron}

\author{A. Pupasov-Maksimov$^a$ and D. Karlovets$^b$}
\address{(a) Universidade Federal de Juiz de Fora, Brasil}
\address{(b) Tomsk State University, Russia}
\ead{tretiykon@yandex.ru,d.karlovets@gmail.com}


\vspace{10pt}
\begin{indented}
\item[]Janyary 2021
\end{indented}

%
\vspace{2pc}

\begin{abstract}
A wide variety of emission processes by electron wave packets with an orbital angular momentum $\ell \hbar$, called the vortex electrons, can be influenced by a nonparaxial contribution due to their intrinsic electric quadrupole moment. We study Smith-Purcell radiation from a conducting grating generated by a vortex electron, described as a generalized Laguerre-Gaussian packet, which has an intrinsic magnetic dipole moment and an electric quadrupole moment. By using a multipole expansion of the electromagnetic field of such an electron, we employ a generalized surface-current method, applicable for a wide range of parameters. The radiated energy contains contributions from the charge, from the magnetic moment, and from the electric quadrupole moment, as well as from their interference. The quadrupole contribution grows as the packet spreads while propagating, and it is enhanced for large $\ell$. In contrast to the linear growth of the radiation intensity from the charge with a number of strips $N$, the quadrupole contribution reveals an $N^3$ dependence, which puts a limit on the maximal grating length for which the radiation losses stay small. We study spectral-angular distributions of the Smith-Purcell radiation both analytically and numerically and demonstrate that the electron's vorticity can give rise to detectable effects for non-relativistic and moderately relativistic electrons. On a practical side, preparing the incoming electron's state in a form of a non-Gaussian packet with a quadrupole moment -- such as the vortex electron, an Airy beam, a Schr\"odinger cat state, and so on -- one can achieve quantum enhancement of the radiation power compared to the classical linear regime. Such an enhancement would be a hallmark of a previously unexplored quantum regime of radiation, in which non-Gaussianity of the packet influences the radiation properties much stronger than the quantum recoil.
\end{abstract}

\maketitle
\section{Introduction}

It was argued that different radiation processes with the vortex electrons carrying orbital angular momentum $\ell\hbar$ with respect to a propagation axis can be investigated using beams of the electron microscopes \cite{bliokh2017theory, ivanov2013detecting}.
For instance, Vavilov-Cherenkov radiation and transition radiation are affected by vortex structure of the electron wave packet \cite{ivanov2016quantum, ivanov2013detecting} and an azimuthal asymmetry of the transition radiation, if detected, would manifest the magnetic moment contribution to the radiation. 
Another radiation process, which we study in the present paper, is the Smith-Purcell (S-P) radiation \cite{smith1953visible} of the vortex electrons. 
Specifically, we investigate how the OAM and the spatial structure of the vortex wave packet influence the radiation characteristics, such as the spectral-angular distributions of the radiated energy. 

\begin{figure}[h!]
 \centering
 \includegraphics[width=.75\linewidth]{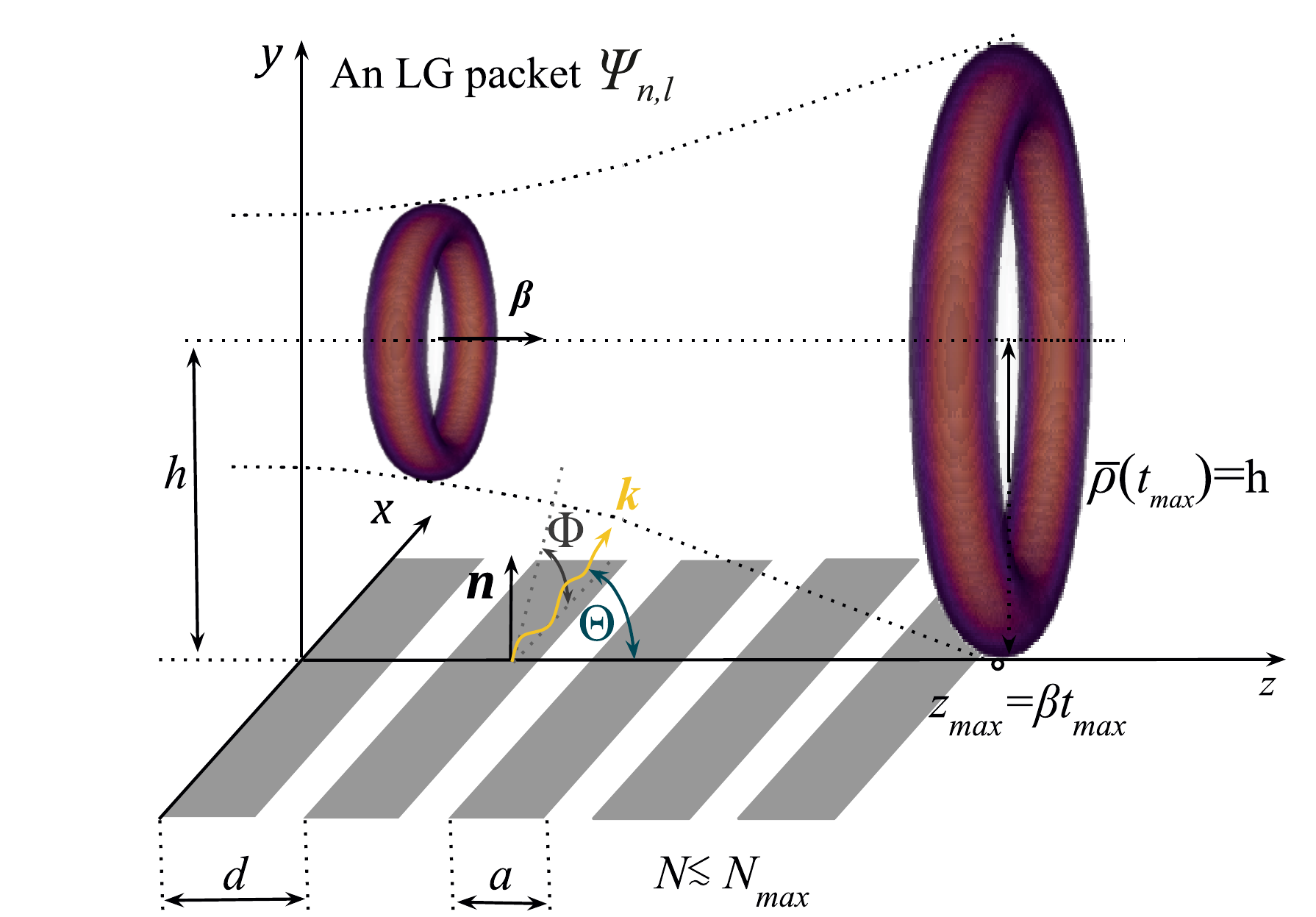}
 \caption{Generation of Smith-Purcell radiation by a Laguerre-Gaussian (LG) packet characterized with a charge $e$, the magnetic moment ${\bm \mu}$, and with the electric quadrupole moment $Q_{\alpha,\beta}$. Two latter quantities are non-vanishing due to intrinsic angular momentum of the vortex electron. The number of the grating strips $N$ cannot be larger than $N_{max}$ due to the spreading.}
 \label{fig:geometry}
\end{figure}

The Smith-Purcell radiation mechanism represents a relatively simple way to generate quasi-monochromatic radiation from charged (electron) bunches passing near a conducting diffraction grating and it has been proved to be useful in developing compact free electron lasers \cite{wachtel1979free,aleinik2004stimulated,wang2007free,bei2008simulation} and high-resolution sensors for the particle beam diagnostics \cite{doucas2001new,kube2003smith,kube2014radiation}. Besides the fundamental interest to the properties of radiation generated by non-Gaussian packets (in particular, by the vortex electrons), there are possible applications in electron microscopy \cite{tolstikhin2019strong} and in acceleration of the vortex electrons via inverse S-P effect \cite{talebi2016schrodinger}. We consider the simplest possible geometry in which the electron wave packet moves above an ideally conducting diffraction grating (see Fig. \ref{fig:geometry}), which is made of $N$ rectangular strips of a width $a$ and with a period $d$. 
The radiation spectrum of a classical charged particle moving with the velocity $\langle u \rangle$ consists of diffraction lines according to the following dispersion relation:
\begin{equation}\label{def:smith-purcell-wave-length}
\lambda_g=\frac{d}{g}\left(\frac{1}{\beta}-\cos \Theta\right),\qquad g=1,2,3,\ldots , 
\end{equation}
where $g$ is the diffraction order,  $\beta$ is the ratio $\beta=\langle u \rangle/c $. The width of the diffracion line is $\Gamma \approx 1/N$. Only $g=1$ is considered below, the radiation wave length is denoted just by $\lambda$ and the frequency $\omega=2\pi/\lambda$.

As we demonstrate hereafter, such a radiation mechanism is more sensitive to the shape of the electron packet than Vavilov-Cherenkov radiation or transition radiation, studied in \cite{ivanov2016quantum, ivanov2013detecting}. A quantum packet always spreads while propagating; however, this does not affect the radiation properties if {\it the radiation formation length} is shorter than the packet's {\it Rayleigh length}, i.e. the distance where the packet doubles its size. A non-relativistic or moderately relativistic electron packet can become significantly wider while moving above a grating with a micrometer or millimeter period, so the Smith-Purcell radiation represents a good tool for studying these effects as the grating can be longer than the Rayleigh length. As we have recently argued, the spreading influences the radiation properties only for wave packets with intrinsic multipole moments \cite{karlovets2020non}\footnote{Which is the case for vortex electrons, while spherically symmetric Gaussian packet do not possess intrinsic multipole moments}.

First, in section \ref{sec:beams-and-moments}, 
recalling basic properties of Laguerre-Gaussian wave packets 
we perform a qualitative analysis to emphasize physics of possible differences from the standard S-P radiation of an ordinary electron. We also establish values of the parameters -- a size of the wave-packet, the orbital angular momentum, velocity, etc. -- that are compatible with our calculation scheme based on a multipole expansion. 

In section \ref{sec:SPR} we calculate the spectral-angular distribution of the radiation applying a method of the generalized surface currents \cite{karlovets2009generalized}, which represents a generalization of the known models put forward by Brownell, et al. \cite{brownell1998spontaneous} and by Potylitsyn, et al. \cite{potylitsyn2000resonant}. Explicit relativistic expressions for the electromagnetic fields produced by a Laguerre-Gaussian wave packet \cite{karlovets2019dynamical} are presented in section \ref{subsec:em-fields}. Calculations of the radiation fields in the wave zone in section \ref{subsec:fourier-integrals} involve standard planar Fourier integrals with respect to $x-z$ coordinates and time. Integration along $z$ direction is tricky because of the wave packet spreading and of the increasing quadrupole moment. To guarantee validity of the multipole expansion when calculating the fields, it is necessary to limit the maximal grating length.     

With analytical expressions at hand, we analyze in section \ref{sec:spectral-angular-distr-SPR}
corrections to the Smith-Purcell radiation of the point charge. In section \ref{subsec:spectral-distr-SPR} we analyze the shape and the position of the spectral line. Our analytical results suggest, that the spreading of the quantum wave packet does not lead to a broadening of the spectral line (in a contrast to the case of a classical spreading beam). Numerical studies of the spectral lines reveal not only an absence of the broadening, but even \textit{a slight narrowing} of the lines due to the charge-quadrupole interference. 

The angular distribution is considered in section \ref{sec:angular-distr-SPR}. The contribution from the magnetic moment results in the azimuthal asymmetry similar to diffraction radiation \cite{ivanov2013detecting}. In section \ref{subsec:coherent-effect} we demonstrate that the quadrupole contribution is dynamically enchanced along the grating. Such a coherent effect can be seen in the nonlinear growth of the radiation intensity with the grating length. At the same time, the maximum of the radiation intensity with respect to the polar angle is shifted towards smaller angles. 

For the currently achieved OAM values of $\ell\sim 1000$ \cite{mafakheri2017realization} (see also \cite{mcmorran2011electron,zhong2019atomistic,zhong2019high}), contributions from both the magnetic moment and the electric quadrupole moment can be, in principle, detected as discussed in the Conclusion. 

Throughout the paper we use the units with $\hbar=c=|e|=1$.


\section{Vortex electrons and multipole moments \label{sec:beams-and-moments}}
\subsection{Laguerre-Gaussian packets and non-paraxial regime \label{subsec:LG-nonparaxial}}
Analogously to optics, there are two main models of the vortex particles -- the Bessel beams and the paraxial Laguerre-Gaussian (LG) packets \cite{bliokh2017theory}.
There is also a non-paraxial generalization of the latter, called the generalized LG packet \cite{karlovets2019dynamical},
\begin{eqnarray}
& \displaystyle \psi_{\ell, n}({\bm r},t) = \sqrt{\frac{n!}{(n + |\ell|)!}} \frac{i^{2n+\ell}}{\pi^{3/4}}\frac{\rho^{|\ell|} \ell^{\frac{2|\ell|+3}{4}}}{(\bar{\rho}(t))^{|\ell| + 3/2}}\ L_{n}^{|\ell|}\left(\frac{\ell \rho^2}{(\bar{\rho}(t))^2}\right) \exp\Big\{-it\frac{\langle p\rangle^2}{2m} + i\langle p\rangle z + i\ell\phi_r - \cr
& \displaystyle - i \left(2n + |\ell| + \frac{3}{2}\right)\arctan (\frac{t}{t_d}) - \frac{\ell( (t_d-it))}{2t_d(\bar{\rho}(t))^2}\,\left(\rho^2 + (z-\langle u\rangle t)^2\right)\Big\},\cr
& \displaystyle \int d^3 r\, |\psi_{\ell, n}({\bm r},t)|^2 = 1,\qquad \rho=\sqrt{x^2+y^2}\,,
\label{LGpsi}
\end{eqnarray}
which represents an exact non-stationary solution to the Schr\"{o}dinger equation and whose centroid propagates along the $z$ axis. The paraxial LG packets and the Bessel beams represent two limiting cases of this model \cite{karlovets2018relativistic}. Below we consider this packet with $n=0$ only. Note that the factor $3/2$ in the Gouy phase in (\ref{LGpsi}) comes about because the packet is localized in a 3-dimensional space (cf. Eq.(61) in \cite{karlovets2018relativistic}).

In what follows, we employ the mean radius of the wave packet $\bar{\rho}_0$ instead of the beam waist $\sigma_{\perp}$,
$$
\sigma_{\perp} = \bar{\rho}_0/\sqrt{|\ell|},\quad \bar{\rho}(t)  =\bar{\rho}_0 \sqrt{1 + t^2/t_d^2}.
$$ 
As explained in \cite{padgett2015divergence}, depending on the experimental conditions one could either fix the beam waist, and so the mean radius $\bar{\rho}_0$ scales as $\sqrt{|\ell|}$, or fix the radius itself. In this paper, we follow the latter approach and treat $\bar{\rho}_0$ as an OAM-independent value. An approach with the fixed beam waist $\sigma_{\perp}$ can easily be recovered when substituting $\bar{\rho}_0 \to \sigma_{\perp}\,\sqrt{|\ell|}$.

Although non-paraxial effects are nearly always too weak to play any noticeable role, the difference between this non-paraxial LG packet and the paraxial one becomes crucial for a moderately relativistic particle with $\beta \lesssim 0.9$. In this regime, which is the most important one for the current study, it is only Eq.(\ref{LGpsi}) that yields correct predictions for observables and is compatible with the general CPT-symmetry \cite{karlovets2019dynamical}. As we argue below, the non-paraxial effects are additionally enhanced when the packet's spreading becomes noticeable and the OAM is large.

The packet spreads and its transverse area is doubled during a diffraction time $t_d$,
\begin{eqnarray}
t_d = \frac{m \bar{\rho}_0^2}{|\ell|} = \frac{t_c}{|\ell|}\, \left (\frac{\bar{\rho}_0}{\lambda_c}\right)^2 \gg t_c,
\label{def:diffraction-time}
\end{eqnarray}
which is large compared to the Compton time scale $t_c = \lambda_c/c \approx 1.3\times 10^{-21}\,\text{sec.}$, $\lambda_c \approx 3.9\times 10^{-11}\, \text{cm}$. When the LG packet moves nearby the grating, the finite spreading time \eqref{def:diffraction-time} puts an upper limit on the possible impact parameter $h$, on the initial mean radius of the packet and on the grating length. Indeed, both the corresponding solution of the Schr\"{o}dinger equation and the multipole expansion have a sense only for as long as $\bar{\rho}(t)<h$ or, alternatively, 
\begin{equation}\label{uneq:diffraction-time}
    \frac{t}{t_d}<\sqrt{\frac{h^2}{\bar{\rho}_0^2}-1}.
\end{equation}
Let $t_{max}=t_d\sqrt{\frac{h^2}{\bar{\rho}_0^2}-1}$ be the time during which the packet spreads to the extent that it touches the grating\footnote{This results in the so-called grating transition radiation, which we do not study in this paper, although the problem in which a part of the electron packet touches the grating and another part does not is definitely interesting to explore.}; then the corresponding number of strips $N_{max}$ is
\begin{eqnarray}
\displaystyle
N_{max}=\beta t_{max}/d=\frac{\beta}{|\ell| }\, \frac{\bar{\rho}_0}{\lambda_c}\,\frac{h}{d}\, \sqrt{1-\frac{\bar{\rho}_0^2}{h^2}}\,.
\label{N-max-expression}
\end{eqnarray}
The geometry implies that $\bar{\rho}_0<h=\bar{\rho}(t_{max})$ or $\bar{\rho}_0\ll h$ for the long grating. In practice, only small diffraction orders of the radiation can be considered, so that $d\sim\beta\lambda$ for the emission angles $\Theta \sim 90^{\circ}$. So an upper limit for the number of strips in this case is
\begin{eqnarray}
& \displaystyle
N_{max}\lesssim \frac{1}{|\ell|}\,\frac{\bar{\rho}_0}{\lambda_c}\,\frac{h}{\lambda}.
\label{N-limitations}
\end{eqnarray}
If this condition is violated, the multipole expansion {\it is no longer applicable}. A rough estimate of the maximal number of strips for $h \approx h_{eff} \sim 0.1 \lambda, \beta \approx 0.5, \bar{\rho}_0 \sim 1$ nm yields
\begin{eqnarray}
& \displaystyle
N_{max}\lesssim \frac{10^3}{|\ell|}.
\label{Nmaxrough}
\end{eqnarray}
So if $N_{max} \gg 1$, then $|\ell| < 10^3$. 

The Laguerre-Gaussian electron packet carries, in addition to the charge, higher multipole moments \cite{karlovets2019intrinsic,silenko2019electric}. 
In particular, at the distances larger than the mean radius of the vortex packet,
\begin{eqnarray}
& \displaystyle
r \gtrsim \bar{\rho}(t),
\label{rho}
\end{eqnarray}
it is sufficient to keep a magnetic dipole moment and an electric quadrupole moment \cite{karlovets2018relativistic, karlovets2019dynamical}, 
\begin{eqnarray}
& \displaystyle {\bm \mu} = {\hat{\bm z}}\,\frac{\ell}{2m} \left(1 - \frac{1}{2}\,\ell^2\, \frac{\lambda_c^2}{\bar{\rho}_0^2}\right),\qquad \, Q_{\alpha\beta}(t) = \left(\bar{\rho}(t)\right)^2\, \text{diag}\{1/2,1/2,-1\}.
\label{Moments}
\end{eqnarray} 
The magnetic moment includes a non-paraxial correction according to Eq.(45) in \cite{karlovets2018relativistic}, which is written for the case $|\ell| \gg 1$. The field of the electron's quadrupole moment originates from a \textit{non-point source} as the quadrupole has a finite width, which is just equal to an rms-radius of the packet.

Note that although an LG packet with $\ell=0$ has a vanishing quadrupole moment, an OAM-less packet with a non-vanishing quadrupole momentum can be easily constructed by making this packet highly asymmetric in shape. Thus our conclusions below can also be applied to an arbitrary non-symmetric wave packet with a non-vanishing quadrupole moment.

When the spreading is essential -- at $t \gtrsim t_d$ -- the inequality (\ref{rho}) can be violated and the multipole expansion cannot be used at all.
Thus, the conventional (paraxial) regime of emission takes place only when the spreading is moderate, $t \lesssim t_d$. Remarkably, the non-paraxial regime of emission favors \textit{moderately large values of the OAM}, in contrast to the enhancement of the magnetic moment contribution for which the OAM should be as large as possible. This is because the quadrupole moment has a finite radius and so the radiation is generated as if the charge were continuously distributed along all the coherence length and not confined to a point within this length \cite{karlovets2020non}.

\subsection{Qualitative analysis and multipole expansion \label{subsec:qualitative-analysis}}




Single-electron regime with a freely propagating packet is realized for low electron currents -- below the so-called start current, which is typically about 1 mA \cite{andrews2005dispersion}.
The charge, the magnetic dipole and the electric quadrupole moments \eqref{Moments} induce surface currents on the grating. These currents, in their turn, generate electric and magnetic fields ${\bm E}_e$, ${\bm E}_\mu$, ${\bm E}_Q$, etc. The total radiation intensity $dW$ includes the multipole radiation intensities as well their mutual interference, which serve as small corrections to the classical radiation from the point charge $dW_{ee}$:
\begin{equation}\label{def:radiation-intensity-interference-complete}
\frac{dW}{d\omega d\Omega}\equiv dW=dW_{ee}+dW_{e\mu}+(dW_{eQ}+dW_{\mu\mu})+(dW_{\mu Q}+dW_{eO})+(dW_{QQ}+dW_{e16p}+dW_{\mu O})+\ldots
\end{equation}
In this paper, we adhere to such a perturbative regime and formally order perturbative corrections following the order of the multipole expansion. The leading order (LO) correction  $dW_{e\mu}$ is given by the charge-magnetic-moment radiation. The next-to-leading (NLO) order corrections include the charge-electric-quadrupole radiation $dW_{eQ}$ and the radiation of the magnetic moment $dW_{\mu\mu}$. 
The next-to-next-to-leading (NNLO) order corrections already include the interference term between the magnetic momentum and electric-quadrupole radiation $dW_{\mu Q}$ and the higher multipole term with the charge-octupole radiation  $dW_{eO}$. The quadrupole-quadrupole radiation 
 $dW_{QQ}$ appears with interference terms from higher multipoles (octupole $dW_{\mu O}$ and 16-pole $dW_{e16p}$).

Moreover, in the overwhelming majority of practical cases it is sufficient to calculate the charge contribution and the following interference terms
\begin{equation}\label{def:radiation-intensity-interference}
dW=dW_{ee}+dW_{e\mu}+dW_{eQ}
\end{equation}
only, while $dW_{\mu \mu}$, $dW_{\mu Q}$, and the higher-order corrections can be safely neglected. We emphasize that it does not mean that there are simple inequalities like $dW_{ee} \gg dW_{e\mu}\gg dW_{eQ} \gg dW_{\mu Q}...$ for all the angles and frequencies. For instance, in the plane perpendicular to the grating, $\Phi = \pi/2$, the term $dW_{e\mu}$ vanishes while $dW_{eQ}$ does not (see below). This makes the region of angles $\Phi \approx \pi/2$ {\it preferable} for detection of the non-paraxial quadrupole effects.

An approach in which the particle trajectory is given holds valid when the quantum recoil $\eta_q$ (ratio between the energy of the emitted photon and the electron's kinetic energy $\varepsilon$) is small compared to the interference corrections $dW_{e\mu}$ and $dW_{eQ}$, 
\begin{eqnarray}
& \displaystyle
\eta_q:= \frac{\omega}{\varepsilon} \ll \frac{dW_{e\mu}}{dW_{ee}},\quad 
\frac{\omega}{\varepsilon} \ll \frac{dW_{eQ}}{dW_{ee}},
\label{def:quasi-classical-restrictions}
\end{eqnarray}
and the energy losses stay small compared to the electron's energy.
We emphasize that the multipole contributions in Eq.(\ref{def:radiation-intensity-interference-complete}) also have a quantum origin as they are due to non-Gaussianity of the wave packet or, in other words, due to its non-constant phase. So the series (\ref{def:radiation-intensity-interference-complete}) is not quasi-classical. On a more fundamental level, there are two types of quantum corrections to the classical radiation of charge \cite{berestetskii1982quantum,bagrov1993theory,akhiezer1990theory,akhiezer1993semiclassical,baier1968processes}: 
\begin{itemize}
    \item Those due to recoil,
    \item Those due to finite coherence length of the emitting particle.
\end{itemize}  
While the quasi-classical methods like an operator method \cite{baier1968processes} and the eikonal method \cite{akhiezer1993semiclassical} neglect the latter effects from the very beginning and take into account the recoil only, here we demonstrate that there is an opposite {\it non-paraxial} regime of emission.

Let us study dimensionless parameters that define multipole corrections to the classical emission of a charge. The magnetic moment contribution 
(see Eqs. \eqref{expr-dWee-SP},\eqref{expr-dWemu-SP}) is proportional to the following ratio:
\begin{eqnarray}
& \displaystyle
\frac{dW_{e\mu}}{dW_{ee}} \sim \eta_{\mu}:=\ell \frac{\lambda_c}{\lambda},
\label{def:emu-parameter}
\end{eqnarray}
which is of the order of $\ell \cdot 10^{-7}$ for $\lambda \sim 1\, \mu$m, and $\ell\cdot 10^{-10}$ for $\lambda\sim 1$ mm.
It is well known that a spin-induced magnetic moment contribution -- the so-called "spin light" \cite{bordovitsyn1995spin} -- and the recoil effects are of the same order of magnitude; thus we neglect both of them in our approach. However, for the vortex electron the magnetic moment contribution \eqref{def:emu-parameter} is $\ell$ times enhanced, which legitimates the calculations of $dW_{e\mu}$ via the multipole expansion \cite{ivanov2013detecting}. Indeed, for the electrons with $\beta \approx 0.4 - 0.8$ and the kinetic energy of $\varepsilon_c \sim 50 - 300$ keV, the small parameter governing the quantum recoil is
\begin{eqnarray}
& \displaystyle
\eta_q = \frac{\omega}{\varepsilon} \sim \frac{1}{m \lambda} \equiv \frac{\lambda_c}{\lambda},  
\label{etaq}
\end{eqnarray}
and $\eta_q\ll \eta_\mu$ yields
$$
|\ell| \gg 1,
$$
while the condition $\eta_\mu^2 \ll \eta_q$ puts \textit{an upper limit} on the OAM value,
\begin{eqnarray}
& \displaystyle
|\ell| \lesssim \ell_{\text{max}} := \sqrt{\frac{\lambda}{\lambda_c}} \sim \eta_q^{-1/2} \sim 10^3 - 10^5,\ \lambda \sim 1\,\mu\text{m} - 1\,\text{mm},  
\label{ineq3}
\end{eqnarray}
and so the contribution $dW_{\mu\mu}$ stays small. 

As shown in section \ref{subsec:radiation-fields}, one can distinguish three different corrections from the charge-quadrupole interference: $dW_{eQ_0}$, $dW_{eQ_1}$, and $dW_{eQ_2}$. Their relative contributions are:
\begin{eqnarray}
&& \displaystyle
\frac{dW_{eQ_0}}{dW_{ee}} \sim \eta_{Q_0}:= \frac{\bar{\rho}_0^2}{h_{\text{eff}}^2} - \text{quasi-classical quadrupole contribution},
\label{def:eQ0-interf-term-parameter}\\
&& \displaystyle
\frac{dW_{eQ_{1}}}{dW_{ee}} \sim \eta_{Q_1}:= \ell^2  \frac{\lambda_c^2}{\bar{\rho}_0^2} - \text{ordinary non-paraxial contribution \cite{karlovets2018relativistic}},
\label{def:eQ1-interf-term-parameter}\\
&& \displaystyle
\frac{dW_{eQ_{2}}}{dW_{ee}} \sim  \eta_{Q_2}:= N^2\, \ell^2 \frac{\lambda_c^2}{\bar{\rho}_0^2}  - \text{dynamically enhanced non-paraxial contribution \cite{karlovets2019dynamical}},
\label{def:eQ2-interf-term-parameter}
\end{eqnarray}
where an effective impact parameter of Smith-Purcell radiation naturally appears 
\begin{equation}
\ h_{\text{eff}} = \frac{\beta\gamma\lambda}{2\pi} = \frac{\beta\gamma}{\omega} \sim 0.1\, \lambda\ \text{for}\ \beta \approx 0.4-0.8.
\label{def:effective-impact-parameter}
\end{equation}
The non-paraxial correction to the magnetic moment in (\ref{Moments}) yields a correction to $\eta_{\mu}$ of the order of $\eta_{\mu}\, \eta_{Q_1}$, which can be safely neglected for our purposes.
 
As seen from \eqref{def:eQ1-interf-term-parameter}, the non-paraxial regime \textit{does not necessarily imply a tight focusing}, $\bar{\rho}_0 \gtrsim \lambda_c$, but it can also be realized when the OAM is large, $\ell \gg 1$, whereas the focusing stays moderate, $\bar{\rho}_0 \gg \lambda_c$. As we fix $\bar{\rho}_0$ and not the beam waist, the parameter (\ref{def:eQ1-interf-term-parameter}) scales as $\ell^2$, which for the electrons with $\bar{\rho}_0 \sim 10$ nm and $\ell \sim 10^3$ yields 
$$
\ell^2\left(\frac{\lambda_c}{\bar{\rho}_0}\right)^2 \sim 10^{-3},
$$
while it is 
$10^{-5}$ for $\bar{\rho}_0 \sim 100$ nm. 

Importantly, these non-paraxial (quadrupole) effects \textit{are dynamically enhanced} when the packet spreading is essential on the radiation formation length, which for Smith-Purcell radiation is defined by the whole length of the grating. Spreading of the packet with the time and distance $\langle z\rangle = t\beta$ leads to growth of the quadrupole moment and a corresponding small dimensionless parameter $\eta_{Q_2}$ is 
$$
t^2/t_d^2 \to N^2
$$ 
times larger than (\ref{def:eQ1-interf-term-parameter}):
\begin{eqnarray}
&& \displaystyle \eta_{Q_2}:= N^2\, \eta_{Q_1} = N^2\, \ell^2\left(\frac{\lambda_c}{\bar{\rho}_0}\right)^2.
\label{param2}
\end{eqnarray}
Thus the large number of strips $N \gg 1$ can lead to \textit{the non-paraxial regime of emission} with 
\begin{eqnarray}
&& \displaystyle \eta_{Q_1} \ll 1,\ \eta_{Q_2} \lesssim 1,
\label{paramineq}
\end{eqnarray}
when the quadrupole contribution becomes noticeable. Somewhat contrary to intuition, these non-paraxial effects {\it get stronger when the packet itself gets wider}, see \eqref{dWeQ2-optimal-rho-l-dependence}.




In the OAM-less case, $n=0$, $\ell=0$, the packet \eqref{LGpsi} turns into the ordinary, spherically symmetric Gaussian packet, which has a vanishing quadrupole  and higher moments. Therefore, its spreading does not lead to such a non-linear enhancement and the Smith-Purcell radiation from this packet in the wave zone coincides with that from a point charge (see also \cite{karlovets2020non}). 
Note that if we fix the beam waist instead and, therefore, $\bar{\rho}_0 \propto \sqrt{|\ell|}$, then
\begin{equation}
\eta_{Q_0} = \mathcal O(\ell),\, \eta_{Q_1} = \mathcal O(\ell),\, \eta_{Q_2} = \mathcal O(\ell N^2).
\label{smalparamestim}
\end{equation}

The dimensionless parameters from the NNL-order corrections $dW_{\mu\mu}$, $dW_{\mu Q_j}$, $dW_{Q_jQ_j}$, $dW_{eO}$ are just products of the leading and NL-order parameters,
\begin{eqnarray}
& \displaystyle
\eta_{\mu\mu}=\eta_\mu^2\,,\quad
\eta_{\mu Q_j}=\eta_{\mu}\eta_{Q_j}\,,\qquad
\eta_{Q_iQ_j}=\eta_{Q_i}\eta_{Q_j},\quad i=0,1,2,\ j=0,1,2.
\label{def:muQiQj-interf-term-parameter}
\end{eqnarray}
The following inequalities are in order
\begin{eqnarray}
& \displaystyle
\eta_q\ll \eta_\mu\,,\quad
\eta_q\ll\eta_{Q_j}\,,\quad
\eta_{\mu\mu}\lesssim \eta_q\,,\quad
\eta_{\mu Q_j}\lesssim \eta_q\,,\quad
\eta_{Q_iQ_j}\lesssim \eta_q,\cr 
& \displaystyle i=0,1,2,\ j=0,1,2,
\label{def:muQiQj-parameter-restrictions}
\end{eqnarray}




For the same beam energies, we have the following estimate for the first quadrupole parameter:
\begin{eqnarray}
& \displaystyle
h_{\text{eff}} \sim 0.1\,\lambda,\quad \eta_{Q_0} \sim 10^2\, \frac{\bar{\rho}_0^2}{\lambda^2}.
\label{relativistic-restrictions-mass-impact}
\end{eqnarray}
According to \eqref{def:muQiQj-parameter-restrictions} the inequalities $\eta_{Q_0}^2\lesssim \eta_q \ll \eta_{Q_0}$ restrict 
the initial rms-radius of the packet as follows:
\begin{eqnarray}
& \displaystyle
\ell_{\text{max}}^{-1} \ll \frac{\bar{\rho}_0}{h_{eff}}\lesssim \ell_{\text{max}}^{-1/2},
\label{rho-bound}
\end{eqnarray}
which yields
\begin{eqnarray}
& \displaystyle
\lambda \sim 1\, \mu\text{m}:\ 0.1\, \text{nm} \ll \bar{\rho}_0 \lesssim 3\, \text{nm},\cr
& \displaystyle
\lambda \sim 1\,\text{mm}:\ 1\, \text{nm} \ll \bar{\rho}_0 \lesssim 300\, \text{nm}.
\label{rho-bound2}
\end{eqnarray}
The packet radius should be smaller than the wavelength of the emitted radiation, which is just a condition of the multipole expansion in the wave zone. 

The inequality $\eta_{Q_1}^2 \lesssim \eta_q\ll \eta_{Q_1}$ defines either a lower bound on $\ell$ or an upper bound for the rms-radius:
\begin{eqnarray}
|\ell|\, \lambda_c\, \ell_{\text{max}}^{1/2} \lesssim \bar{\rho}_0 \ll |\ell|\, \lambda_c\, \ell_{\text{max}},
\label{lower-ell-bound}
\end{eqnarray}
which yields
\begin{eqnarray}
& \displaystyle
0.1-1\, \text{nm} \lesssim \bar{\rho}_0 \ll 3\, \text{nm} - 30\, \text{nm},\ \ell \sim 10-100 < \ell_{\text{max}} \sim 10^3,\ \lambda \sim 1\, \mu\text{m},\cr
& \displaystyle 1 - 100\, \text{nm} \lesssim \bar{\rho}_0 \ll 0.3 - 300\, \mu\text{m},\ \ell \sim 10-10^{4} < \ell_{\text{max}} \sim 10^5,\ \lambda \sim 1\, \text{mm}.
\label{lower-ell-bound2}
\end{eqnarray}
This is compatible with (\ref{rho-bound2}) provided that the OAM is at least $\ell\sim 100$.


Finally, the restrictions for the number of strips $N$ can be derived from the inequality $\eta_{Q_2}^2 \lesssim \eta_q \ll \eta_{Q_2}$:
\begin{eqnarray}
& \displaystyle
 \frac{\bar{\rho}_0}{|\ell|\ell_{\text{max}}\lambda_c} \ll N \lesssim N_{max} :=  \frac{\bar{\rho}_0}{|\ell|\ell_{\text{max}}^{1/2}\lambda_c},
\label{Nsbound}
\end{eqnarray}
where the ratio $\bar{\rho}_0/(|\ell|\ell_{\text{max}}\lambda_c)$ itself must be less than unity according to (\ref{lower-ell-bound}). So, the smallest value of $N$ can well be $1$. 

Let us estimate the largest possible number of strips for which our conditions hold.
For an optical or infrared photon, $\lambda \sim 1\, \mu$m, $\ell \sim 10^2$, and $\bar{\rho}_0 \sim 1\, \text{nm}-3\,\text{nm}$ (according to above findings), we get
$$
N_{max} \sim 3.
$$
So the grating must be really short. For a THz photon with $\lambda \sim 1$ mm, $\ell \sim 10^2$, and $\bar{\rho}_0 \sim 100\, \text{nm}-300\,\text{nm}$ we have
$$
N_{max} \sim 30,
$$
or the same number of $N_{max} \sim 3$ for $\ell \sim 10^3$. These inequalities specify the rough estimate (\ref{Nmaxrough}). For Smith-Purcell radiation, the large number of strips provides a narrow emission line, so the optimal OAM value is therefore 
\begin{eqnarray}
& \displaystyle
\ell \sim 10^2-10^3,
\label{ellopt}
\end{eqnarray}
and the optimal grating period, which defines the radiation wavelength as $d\sim\lambda$, is 
\begin{eqnarray}
& \displaystyle
d \sim 10 - 1000\, \mu\text{m}.
\label{dopt}
\end{eqnarray}
For the largest wavelength, the maximal grating length for which the higher-multipole corrections can be neglected and the radiation losses stay small is of the order of 3 cm.

Importantly, the maximal grating length $N_{max} d$ \textit{is much larger} than the Rayleigh length $z_R$ of the packet,
\begin{eqnarray}
& \displaystyle
z_R = \beta t_d = \beta \frac{\lambda_c}{|\ell|}\, \left (\frac{\bar{\rho}_0}{\lambda_c}\right)^2,
\label{zR}
\end{eqnarray}
which is of the order of $0.1\, \mu\text{m}$ for $\lambda\sim 1\,\mu\text{m}$ and the same parameters as above, or $z_R \sim 1$ mm for $\lambda\sim 1$ mm and $\ell\sim 10^2$.

Summarizing, one can choose two baseline parameter sets:
\begin{itemize}
    \item 
(IR):\ $\lambda \sim 1\, \mu\text{m}$, $\bar{\rho}_0=0.5-3\, \text{nm}$, $\ell \sim 100$, $N \lesssim 10$,   
    \item
(THz):\ $\lambda \sim 1\, \text{mm}$, $\bar{\rho}_0=10-300\, \text{nm}$, $\ell \sim 10^2-10^3$, $N \lesssim 100$.
\end{itemize}
As has been already noted, in practice the corresponding inequalities and the subsequent requirements can often be relaxed, as the ratios like $dW_{e\mu}/dW_{ee}$ are generally functions of angles and frequency. For instance, the requirement $\eta_q \ll \eta_{\mu}$ does not have a sense in a vicinity of $\Phi=\pi/2$ as $dW_{e\mu}$ vanishes at this angle. Finally, note that typical widths of the electron packets after the emission at a cathode vary from several Angstrom to a few nm, depending on the cathode \cite{cho2004quantitative, ehberger2015highly}, which meets our requirements. 

\section{Smith-Purcell radiation via generalized surface currents \label{sec:SPR}}
\subsection{Surface currents and radiation field \label{subsec:radiation-fields}}

Following the generalized surface current model developed in Ref.\cite{karlovets2009generalized} we express the current density induced by the incident electromagnetic field of the electron on the surface of an ideally conducting grating as a vector product of ${\bm E}$, the normal to the surface ${\bm n}$ and the unit vector to a distant point 
$$
{\bm e}_0=\frac{{\bm r}_0}{|{\bm r}_0|}=\left(\sin \Theta \cos \Phi,\sin \Theta \sin \Phi,\cos \Theta \right),
$$
\begin{equation}
\label{def:generalized-current}
{\bm j}(w)=\frac{1}{2\pi}\, {\bm e}_0 \times \left[\,{\bm n} \times {\bm E}(w)\right].
\end{equation}
This expression is suitable for calculating the radiated energy in the far-field only, as one should generally have a curl instead of ${\bm e}_0$ and the induced current should not depend on the observer's disposition. 

Unlike the surface current density used in the theory of diffraction of plane waves, this one has all three components, including the component perpendicular to the grating surface. This normal component comes about ultimately because the incident electric field has also all three components, unlike the plane wave. For the ultrarelativistic energies, the normal component of the surface current can be safely neglected and in this case the generalize surface current model completely coincides \cite{karlovets2011theory} with the well-known approach by Brownell, et al. \cite{brownell1998spontaneous}.
The latter model was successfully tested, for instance, in experiment \cite{blackmore2009first} conducted with a $28.5$ GeV electron beam. However for the moderate electron energies, needed for observation of the effects we discuss in this work, the normal component of the surface current is crucially important, which is why we employ the more general model of Ref.\cite{karlovets2009generalized}.

To calculate the radiation fields at large distances we use Eq.(28) from \cite{karlovets2009generalized} 
\begin{equation}
{\bm E}^R\approx \frac{i\omega {\rm e}^{i kr_0}}{2\pi r_0}\int {\bm e}_0  \times \left[\,{\bm n} \times {\bm E} (k_x,y,z,\omega) \right]{\rm e}^{-ik_z z}dz\,,
\end{equation}
where the integration is performed along the periodic grating.

\subsection{Electromagnetic field of a vortex electron} \label{subsec:em-fields}

In Appendix we calculate explicit expressions for the electromagnetic fields produced by a vortex electron \cite{karlovets2019dynamical}  
in the cartesian coordinates.
One can separate the field into the contributions of the charge, of the magnetic moment, 
and of the quadrupole moment as follows:
\begin{flalign}
\label{Eelabsep}
& {\bm E}({\bm r}, t)  = {\bm E}_e({\bm r}, t) + {\bm E}_{\mu}({\bm r}, t) + {\bm E}_Q({\bm r}, t),\cr
& {\bm E}_e({\bm r}, t)  = \frac{1}{R^3} \{\gamma{\bm \rho}, R_z\},\\
\label{Emulabsep}
& {\bm E}_{\mu}({\bm r}, t)  = \frac{\ell}{2m}\frac{3\beta\gamma}{R^5} R_z\{y, -x, 0\},\\
& {\bm E}_{Q}({\bm r}, t) = \frac{\gamma}{4R^3} {\bm \rho}\Bigg (3\frac{\bar{\rho}_0^2}{R^2} \left(1 - 5\frac{R_z^2}{R^2}\right) + \ell^2 \left(\frac{\lambda_c}{\bar{\rho}_0}\right)^2 \Big[3 \frac{T_z^2}{R^2}\left(1 - 5\frac{R_z^2}{R^2}\right) + 3 \frac{R_z^2}{R^2} - 6\beta \frac{R_z T_z}{R^2} -  1\Big]\Bigg) + \cr 
& \frac{\gamma}{4R^3}({\bm z} - {\bm \beta}t) \Bigg (3\frac{\bar{\rho}_0^2}{R^2} \left(3 - 5\frac{R_z^2}{R^2}\right) + \ell^2 \left(\frac{\lambda_c}{\bar{\rho}_0}\right)^2 \Big[ \frac{T_z^2}{R^2}\left(3 - 5\frac{R_z^2}{R^2}\right)+ 3 \frac{R_z^2}{R^2} - 1\Big]\Bigg).
\label{EQlabsep}
\end{flalign}
where following notations are used: ${\bm \beta} = \left(0, 0, \beta\right),\, {\bm z} = \left(0, 0, z\right)$,
\begin{align}
 {\bm \rho} = \{x,y\},\quad R_z:=\gamma(z - \beta t),\quad T_z:=\gamma(t-\beta z),\cr
 {\bm R} = \{{\bm \rho}, \gamma (z - \beta t)\} \,,\quad \gamma=(1-\beta^2)^{-1/2}
\label{notations-R-Rz-Tz}
\end{align} 
We omit the magnetic fields, as to calculate the surface current below we need the electric field only.
In the problem of Smith-Purcell radiation, the grating is supposed to be very long in the transverse direction, so we need the Fourier transform of these fields.

\subsection{Fourier transform of the fields \label{subsec:fourier-integrals}}

When calculating the Fourier transform of the electric fields produced by the wave-packet 
$$
{\bm E}(q_x, y, z,\omega) = \int\limits dxdt\, {\bm E}({\bm r},t) e^{i\omega t - iq_x x}
$$
the following integral and its derivatives appear
\begin{equation}\label{def:master-integrals}
I_\nu(q_x,y,z,\omega)=
\int\limits_{-\infty}^{\infty}dt \int\limits_{-\infty}^{\infty}dx \frac{{\rm e}^{i(\omega t-q_x x)}}{\left(x^2+y^2+\gamma^2(z-\beta t)^2\right)^{(\nu/2)}}\,,\qquad \nu=3,5,7\,.
\end{equation}
We consider 3 master-integrals to reduce a number of derivatives with respect to parameters, although $I_3$ alone would be enough to calculate the Fourier transform of all the terms. 
The master integrals read
\begin{eqnarray}
&& \displaystyle \label{master-integral-I3}
I_3(q_x,y,z,\omega)=\frac{2\pi}{\gamma\beta}\exp\left(\frac{i\omega z}{\beta}-\mu|y|\right)\frac{1}{|y|},\\
&& \displaystyle \label{master-integral-I5}
I_5(q_x,y,z,\omega)=\frac{2\pi}{\gamma\beta}\exp\left(\frac{i\omega z}{\beta}-\mu|y|\right)\frac{(1+\mu |y|)}{3|y|^3},\\
&& \displaystyle \label{master-integral-I7}
I_7(q_x,y,z,\omega)=\frac{2\pi}{\gamma\beta}\exp\left(\frac{i\omega z}{\beta}-\mu|y|\right)\frac{(3+3\mu|y|+\mu^2|y|^2)}{15|y|^5}\,,
\end{eqnarray}
where $\mu=\sqrt{\frac{\omega^2}{\gamma^2\beta^2}+q_x^2}$, $\nu=3,5,7$. 

All the rest can be obtained by taking derivatives of the corresponding master integral either over $t$ or $x$
\begin{equation}\label{fourier-integrals-by-master-integralss}
I_{\nu,t^s,x^p}=\int\limits_{-\infty}^{\infty}dt \int\limits_{-\infty}^{\infty}dx \frac{t^sx^p{\rm e}^{i(\omega t-q_x x)}}{\left(x^2+y^2+\gamma^2(z-\beta t)^2\right)^{(\nu/2)}}=
i^{p-s}\partial_\omega^s\partial_{q_x}^p I_\nu(q_x,y,z,\omega)\,.
\end{equation}
Note that only $p=0$ ($y$ and $z$ components of electric field) and $p=1$ ($x$ component) cases are required.
In particular, electric fields from the charge and the magnetic momentum read
\begin{align}
&
{\bm E}_e(q_x, y, z,\omega) = \gamma 
\left( i \partial_{q_x},y, \left(z+i \beta \partial_\omega\right) \right)\,I_3(q_x, y, z,\omega)\,,\label{Ee-by-master-int}\\
&
{\bm E}_\mu(q_x, y, z,\omega) = \frac{3\ell \beta\gamma^2}{2m} 
\left(z+i \beta \partial_\omega\right) \left(y,-i \partial_{q_x},0 \right)\,I_5(q_x, y, z,\omega)\,,\label{Emu-by-master-int}
\end{align}
which after the differentiation reads
\begin{align}
&  {\bm E}_e(q_x, y, z,\omega) = \frac{2\pi}{\beta}  \left(-iq_x, \text{sgn}(y) \sqrt{\left(\frac{\omega}{\beta \gamma}\right)^2 + q_x^2}, -i\frac{\omega}{\beta\gamma^2}\right)\,\frac{\exp\left\{iz\frac{\omega}{\beta} - |y| \sqrt{\left(\frac{\omega}{\beta \gamma}\right)^2 + q_x^2}\right\}}{\sqrt{\left(\frac{\omega}{\beta \gamma}\right)^2 + q_x^2}},\cr
&
{\bm E}_{\mu}(q_x, y, z,\omega) = -\frac{\ell}{2m}\frac{i2\pi\omega}{\beta\gamma} 
\left(\text{sgn}(y) \sqrt{\left(\frac{\omega}{\beta \gamma}\right)^2 + q_x^2}, iq_x, 0\right)\frac{\exp\left\{iz\frac{\omega}{\beta} - |y| \sqrt{\left(\frac{\omega}{\beta \gamma}\right)^2 + q_x^2}\right\}}{\sqrt{\left(\frac{\omega}{\beta \gamma}\right)^2 + q_x^2}},
\label{Eqx}
\end{align}
where $q^2 = q_0^2-{\bm q}^2 \ne 0$ and for the electron packet whose center is at the distance $h$ from the grating one needs to substitute $|y| \rightarrow |y-h|$.

Technically, the Fourier transform of the quadrupole fields follows the same line. Starting from the formula \eqref{EQlabsep}, one should substitute $x$ and $t$ variables in the numerator by the differential operators  $x\to i \partial_{q_x}$, $t\to -i \partial_\omega$, acting on the master integrals defined by the denominators, $R^{-\nu/2}\to I_\nu$. Resulting expressions are calculated with the aid of computer algebra and can be found in the public repository \cite{pupasov2019git}. Here we only discuss the general structure of the corresponding expressions. Consider the Fourier transform of a term where $R_z=\gamma(z-\beta t)$ enters the numerator  
$$
\int\limits dxdt\, f(x,y)\frac{R_z^{n}}{R^{\frac{\nu}{2}}} e^{i\omega t - iq_x x}=
\gamma^{n}f(i\partial_{q_x},y)(z+i\beta\partial_\omega)^{n-1}(z+i\beta\partial_\omega) {\rm e}^{i\omega z/\beta}I_f(q_x,y,\omega)\,.
$$
The commutator $[i\beta\partial_\omega, {\rm e}^{i\omega z/\beta}]=-z{\rm e}^{i\omega z/\beta}$ implies that $(z+i\beta\partial_\omega) {\rm e}^{i\omega z/\beta}I_f(q_x,y,\omega)={\rm e}^{i\omega z/\beta}(i\beta\partial_\omega) I_f(q_x,y,\omega)$ and 
we get the integral where  
$z$-variable enters only the exponential factor
\begin{equation}\label{FT-Q-term}
\int\limits dxdt\, f(x,y)\frac{R_z^n}{R^{\frac{\nu}{2}}} e^{i\omega t - iq_x x}=\gamma^n{\rm e}^{i\omega z/\beta} f(i\partial_{q_x},y)(i\beta\partial_\omega)^{n} I_f(q_x,y,\omega)\,.
\end{equation}
Therefore we express $t$ using $z$ and $R_z$,
$$t-\beta z=-\frac{1}{\beta}(z-\beta t)+\frac{z}{\beta \gamma^2}=-\frac{R_z}{\beta\gamma}+\frac{z}{\beta \gamma^2}$$
and rewrite \eqref{EQlabsep} as a sum of \eqref{FT-Q-term}-like terms.   
As a result, the Fourier transform of the quadrupole fields has the following factorized structure:
\begin{eqnarray}
&& \displaystyle {\bm E}_Q(q_x, y, z,\omega) = 
\exp \left( i z\frac{\omega}{\beta} \right)
\left({\bm E}_{Q_0}(q_x, y,\omega)+{\bm E}_{Q_1}(q_x, y,\omega)z+{\bm E}_{Q_2}(q_x, y,\omega)z^2\right),
\label{E-quadrupole-z-polynomial}
\end{eqnarray}
where a $z$-dependent plane-wave is multiplied by a second order polynomial in $z$-variable with coefficients being some functions.
Note that the constant term of the polynomial has the leading term proportional to the charge contribution
$$
\exp \left( i z\frac{\omega}{\beta} \right)
{\bm E}_{Q_0}(q_x, y,\omega)=\frac{\ell^2 \lambda_c^2}{ \bar{\rho}_0^2}{\bm E}_e(q_x, y,z,\omega)+\ldots
$$
The charge and the magnetic dipole contributions depend on $z$ due to the $z$-dependent plane-wave factor only. That is the Fourier transform of the total electric field  
has the same structure as in \eqref{E-quadrupole-z-polynomial},
\begin{eqnarray}
&& \displaystyle {\bm E}(q_x, y, z,\omega) = 
\exp \left( i z\frac{\omega}{\beta} \right)
\left({\bm E}_0(q_x, y,\omega)+{\bm E}_{Q_1}(q_x, y,\omega)z+{\bm E}_{Q_2}(q_x, y,\omega)z^2\right),
\label{E-total-z-polynomial}
\end{eqnarray}
where ${\bm E}_0(q_x, y,\omega)={\bm E}_e(q_x, y,\omega)+{\bm E}_\mu(q_x, y,\omega)+{\bm E}_{Q_0}(q_x, y,\omega)$. The terms linear and quadratic in z contain the quadrupole contribution only and represent the non-paraxial contributions mentioned earlier. We will use this structure in the next section to perform integration along the grating. 

The surface current
$$
{\bm j}=\frac{1}{2\pi}\,\left(-E_x e_{0y},\, E_x e_{0x}+E_z e_{0z},\, -E_z e_{0y}\right)\,,
$$
inherits the structure of Eq.\eqref{E-total-z-polynomial} 
\begin{equation}
\label{def:generalized-current-structure}
{\bm j}(k_x, y, z,\omega) = 
\exp \left( i z\frac{\omega}{\beta} \right)
\left({\bm j}_0(k_x, y,\omega)+{\bm j}_{Q_1}(k_x, y,\omega)\,z+{\bm j}_{Q_2}(k_x, y,\omega)\,z^2\right),
\end{equation}
where the first term contains all types of the contributions, ${\bm j}_{0}(k_x, y,\omega)={\bm j}_e(k_x, y,\omega)+{\bm j}_\mu(k_x, y,\omega)+{\bm j}_{Q_0}(k_x, y,\omega)$, while the next terms are related to the quadrupole contribution only. Note that here ${\bm k} = \omega\, {\bm r}_0/r_0$ is an on-mass-shell wave vector, $k^2 = \omega^2 - {\bm k}^2 = 0$.

Integrating with respect to $z$-coordinate along the periodic grating we get  
\begin{equation}
\int\limits_0^{N d}dz\left({\bm j}_{0}+{\bm j}_{Q_1}\,z+{\bm j}_{Q_2}\,z^2\right)\exp \left( i z \left(\frac{w}{\beta} - k_z\right)\right)=
\label{integral-for-radiation-field}    
\end{equation}
$$
=\left({\bm j}_{0}+{\bm j}_{Q_1}(i\partial_{k_z})+{\bm j}_{Q_2}(i\partial_{k_z})^2\right)F(\Theta_1)\,,
$$
where 
\begin{align}\label{def:gratting-form-factor}
    & 
    F(\Theta_1)=\sum_{j=0}^{N}\int\limits_{jd}^{jd+a}dz \exp \left( i z \left(\frac{\omega}{\beta} - k_z\right)\right)=
\frac{2\sin(\frac{a\Theta_1}{2})}{\Theta_1}\frac{\sin\left(\frac{N d\Theta_1}{2}\right)}{\sin\left(\frac{d\Theta_1}{2}\right)}
\exp\left(\frac{i\Theta_1}{2}(a + (N - 1)d)\right),
\end{align}
and we denote 
$$
\Theta_1 = \frac{\omega}{\beta} - k_z.
$$ 
Here $a$ is a strip width (see Fig.1). Note that $\partial_{k_z}=-\partial_{\Theta_1}$, and to write the resulting formulas in a compact form we will use the following notations: $$
\partial_{\Theta_1}F(\Theta_1)=F'(\Theta_1),\ \partial^j_{\Theta_1}F(\Theta_1)=F^{(j)}(\Theta_1).
$$ 

A standard interference factor due to diffraction on a grating is $|F|^2$. As can be seen, the radiation from the charge and the magnetic moment is modulated by the standard interference factor $|F|^2$, while the interference of the charge with the quadrupole involves $F$ and its derivatives. As a result, the non-symmetric shape of the electron packet results in a small\textit{ modification of the Smith-Purcell dispersion relation}, Eq.(1).

\section{Multipole corrections to the spectral-angular distribution of the Smith-Purcell radiation from the LG-wave packet \label{sec:spectral-angular-distr-SPR}}
\subsection{Spectral distribution of the Smith-Purcell radiation from the LG-wave packet \label{subsec:spectral-distr-SPR}}

The distribution of the radiated energy over the frequencies and angles,
\begin{equation}
\frac{d^2W}{d\omega d\Omega} = r_0^2|{\bm E}^R|\,,
\end{equation}
represents a sum of the following terms (cf. Eq.(\ref{def:radiation-intensity-interference})):
\begin{align}
& \label{expr-dWee-by-FF}
dW_{ee}= \frac{\omega^2}{4\pi^2}|{\bm j}_e|^2|F(\Theta_1)|^2\,,\\
& \label{expr-dWemu-by-FF}
dW_{e\mu}=\frac{\omega^2}{4\pi^2}\left[{\bm j}_e{\bm j}_\mu^*+{\bm j}_e^*{\bm j}_\mu\right]|F(\Theta_1)|^2\,,\\
& \label{expr-dWeQ0-by-FF}
dW_{eQ_0}=\frac{\omega^2}{4\pi^2}\left[{\bm j}_e{\bm j}_{Q_0}^*+{\bm j}_e^*{\bm j}_{Q_0}\right]|F(\Theta_1)|^2\,,\\
&  \label{expr-dWeQ1-by-FF}
dW_{eQ_1}=\frac{i\omega^2}{4\pi^2}\left[i{\bm j}_e{\bm j}_{Q_1}^*F(\Theta_1)F'(\Theta_1)^*-{\bm j}_e^*{\bm j}_{Q_1}F(\Theta_1)^*F'(\Theta_1)\right]
\,,\\
& \label{expr-dWeQ2-by-FF}
dW_{eQ_2}=-\frac{\omega^2}{4\pi^2}\left[{\bm j}_e{\bm j}_{Q_2}^*F(\Theta_1)F''(\Theta_1)^*+{\bm j}_e^*{\bm j}_{Q_2}F(\Theta_1)^*F''(\Theta_1)\right]
\,,\\
& \label{expr-dWmuQ-by-FF}
dW_{\mu Q_j}=-\frac{(-i)^j\omega^2}{4\pi^2}\left[{\bm j}_\mu{\bm j}_{Q_j}^*F(\Theta_1)F^{(j)}(\Theta_1)^*-{\bm j}_\mu^*{\bm j}_{Q_j}F(\Theta_1)^*F^{(j)}(\Theta_1)\right]
\,,\\
& \label{expr-dWmumu-by-FF}
dW_{\mu \mu}=\frac{\omega^2}{4\pi^2}|{\bm j}_\mu|^2|F(\Theta_1)|^2\,,\\
& \label{expr-dWQQ-by-FF}
dW_{Q_jQ_k}=\frac{(-i)^{j+k}\omega^2}{4\pi^2}\left[(-1)^k{\bm j}_{Q_j}{\bm j}_{Q_k}^*F^{(j)}(\Theta_1)F^{(k)}(\Theta_1)^*+(-1)^j{\bm j}_{Q_j}^*{\bm j}_{Q_k}F^{(j)}(\Theta_1)^*F^{(k)}(\Theta_1)\right].
\end{align}
From here, the charge radiation and the charge-dipole interference can be explicitly calculated
\begin{eqnarray}\label{expr-dWee}
&& \displaystyle
\frac{d^2W_{ee}}{d\omega d\Omega}=
\exp\left(-\frac{2y}{h_{\text{eff}}}\sqrt{1+\beta^2\gamma^2 \cos^2\Phi \sin^2\Theta}\right)\cr
&& \displaystyle
\times\frac{\cos^2\Theta+2\beta\gamma^2\cos^2\Phi\cos\Theta\sin^2\Theta+\sin^2\Phi\sin^2\Theta+\beta^2\gamma^4\cos^2\Phi\sin^4\Theta}{\gamma^2(1-\beta \cos \Theta)^2(1+\beta^2\gamma^2 \cos^2\Phi \sin^2\Theta)}\,|F|^2\,,\\
\label{expr-dWemu}
&& \displaystyle
\frac{d^2W_{e\mu}}{d\omega d\Omega}=
\frac{\ell}{m} \exp \left(-\frac{2y}{h_{\text{eff}}}\sqrt{1+\beta^2 \gamma^2\cos^2\Phi \sin^2\Theta}\right)\cr
&& \displaystyle
\times\frac{ \omega \cos\Phi \sin \Theta  \left(\beta\gamma^2 \sin^2\Theta+\cos\Theta\right) }{ \gamma ^2  (1-\beta \cos \Theta)^2\sqrt{1+\beta^2 \gamma^2
   \cos^2\Phi \sin^2\Theta}}\,|F|^2\,.
\end{eqnarray}
The charge contribution reproduces\footnote{In our coordinate system $\Phi=0$ corresponds to the $x$ axis on the grating plane (see Fig.1), thus our Eq.\eqref{expr-dWee} turns into Eq. (56) of \cite{karlovets2011theory} after a substitution $\Phi\to\phi+\pi/2$.} Eq.(56) of \cite{karlovets2011theory}. The charge-dipole interference results in an \textit{azimuthal asymmetry} arising from $\cos\Phi$, analogously to other types of polarization radiation \cite{ivanov2013polarization}. As noted earlier, the $e\mu$ contribution vanishes at $\Phi=\pi/2$.

\begin{figure}[h!]
 \centering
 \includegraphics[width=.75\linewidth]{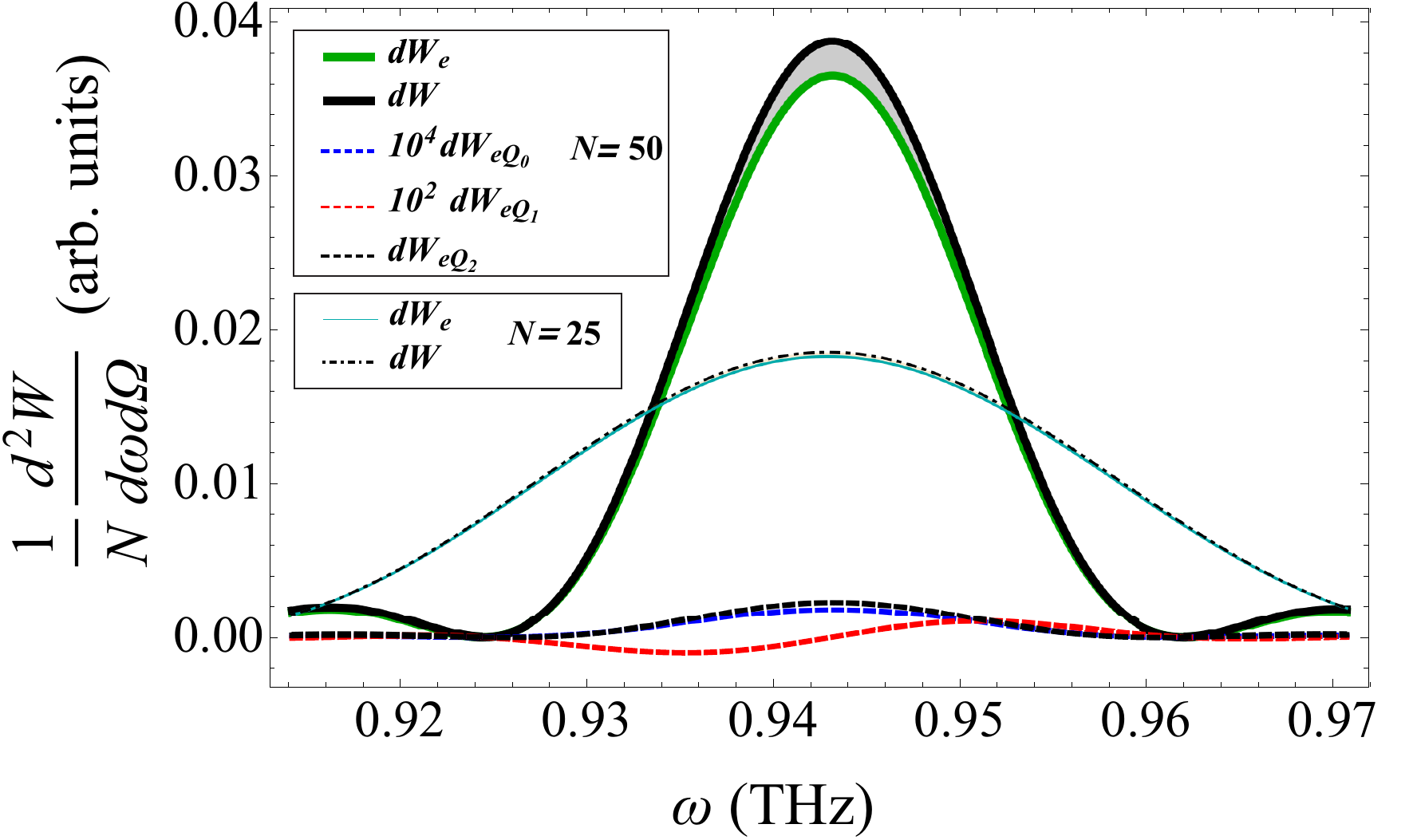}
 \caption{Comparison of the radiation spectrum of an ordinary electron and of a vortex electron packet ($\bar{\rho }_0=300 \text{nm}$, $\ell=1000$) for two gratings with $N=25$ and $N=50$. Quadrupole corrections $dW_{eQ_j}$ are shown for $N=50$ only. Radiation intensities are normalized per 1 strip, the zenith direction perpendicular to the grating plane
 $\Theta=\Phi=\frac{\pi}{2}$ is considered. Difference of the full radiation intensity and the charge one is shown by filling between the corresponding curves.
 The grating period $d=1\,$ mm, $\beta=0.5$, $a=d/2$.}
 \label{fig:spectral-curve-1mm}
\end{figure}

\begin{figure}[h!]
        \includegraphics[width=.99\linewidth]{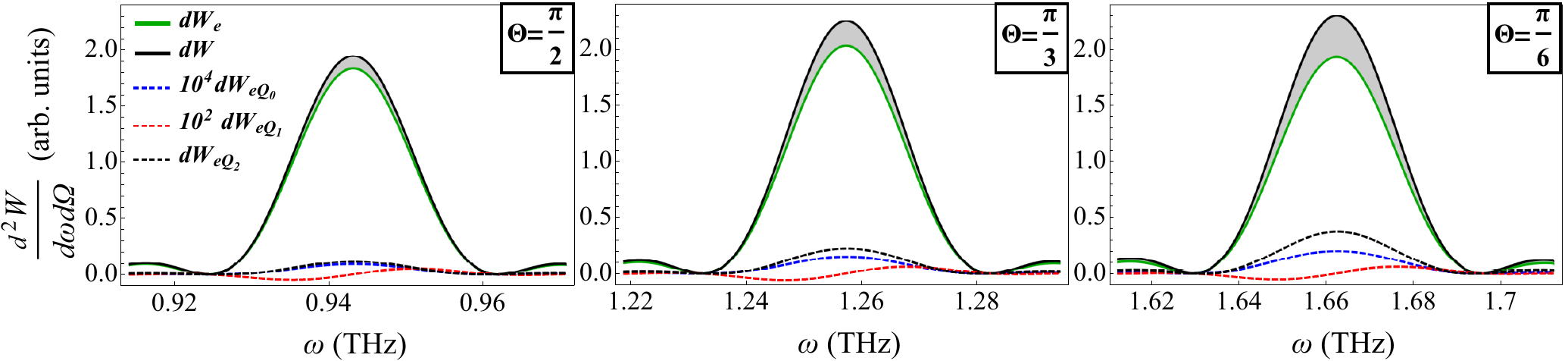}
           \caption{Radiation intensity at different polar angles (black solid line) and contributions from the charge (green solid line), and the electric quadrupole moment (black, red and blue dashed lines) with the following parameters: $\beta=0.5$, $d=1 \text{mm}$, $a=d/2$, $\bar{\rho }_0=300 \text{nm}$, $\ell=1000$, $N=50$, $\Phi=\frac{\pi}{2}$ .}
    \label{fig:3polar-angles}
\end{figure}

The quadrupole contributions from ${\bm j}_{Q_1}$ and ${\bm j}_{Q_2}$ are defined by real parts of the product of currents, by the form factor $F$ and its derivatives. Nevertheless, explicit calculations show \footnote{See the code in the public repository \cite{pupasov2019git}}  that these terms have also a
factorized structure 
\begin{equation}\label{expr-dWeQ12-factorized-structure}
dW_{eQ_j}=\exp \left(-\frac{2y}{h_{\text{eff}}}\sqrt{1+\beta^2 \gamma^2\cos^2\Phi \sin^2\Theta}\right)
P_{eQ_j}(k_x,y,\omega)F_{eQ_j}(\omega,k_z).
\end{equation}
Here, the functions $P_{eQ_j}(k_x,y,\omega)$ define the angular distributions and $F_{eQ_j}(\omega,k_z)$ determine positions of the spectral lines and their width (therefore we will call $F_{eQ_j}(\omega,k_z)$ {\it a spectral factor}) .
In figure \ref{fig:spectral-curve-1mm} we compare the radiation intensities normalized per 1 strip from gratings with $N=25$ and $N=50$ strips.
The spectral curves of $dW_{ee}$, $dW_{eQ_0}$ and $dW_{eQ_2}$ have a similar shape and position; up to the sign this is also the case for $dW_{e\mu}$, which is zero at $\Phi=\pi/2$ by the symmetry considerations. The contribution $dW_{eQ_1}$ leads to a shift of the spectral line, but its amplitude is rather small (a factor of $10^2$ is used in Fig.\ref{fig:spectral-curve-1mm}) and this shift is almost unobservable. A nonlinear amplification of the quadrupole contribution to the radiation intensity is clearly seen in figure \ref{fig:spectral-curve-1mm}. This effect becomes stronger for the radiation in the forward direction (see figure \ref{fig:3polar-angles}).


Figures \ref{fig:spectral-curve-1mm} and \ref{fig:3polar-angles} correspond to the case when both number of strips $N=50$ and OAM, $\ell=1000$ are close to maximal values estimated in section \ref{subsec:qualitative-analysis}. Table \ref{tab:parameters-1mm} gives the corresponding dimensionless parameters.  Note that two higher order terms $\eta_{Q_{12}}$ and $\eta_{Q_{22}}$ also surpass quantum recoil in this case. Moreover, $\eta_{Q_{22}}$, being two orders of magnitude smaller than the leading correction $\eta_{Q_2}$, becomes more important than $\eta_{Q_0}$ and $\eta_{Q_1}$ corrections. This means that within our perturbative method,  only charge and charge-quadrupole $\eta_{Q_2}$ contributions should be computed, while all the rest corrections can be considered as next-to-leading order corrections which are at least two orders of magnitude smaller (we plot the corresponding curves in figures \ref{fig:spectral-curve-1mm} and \ref{fig:3polar-angles} just to demonstrate their shapes).    
\begin{table}
    \begin{tabular}{c|c|c|c|c|c}
 $\eta_q = \omega/\varepsilon$              & $\eta _{\mu}=\ell\lambda_c/\lambda$       & $\eta _{\text{Q0}}=\bar{\rho}_0^2/h_{\text{eff}}^2$    & $\eta _{\text{Q1}} = \ell^2\lambda_c^2/\bar{\rho}_0^2$  & $\eta _{\text{Q2}}=N^2\,\ell^2\lambda_c^2/\bar{\rho}_0^2$   & $N$ \\
 \hline
$ 1.95\times 10^{-10}$ & $ 1.95\times 10^{-7}$ & $2.67\times 10^{-6}$ & $1.69\times 10^{-6}$ & ${\bf 4.22\times 10^{-3}}$ &  $50$ 
   \\
 \hline
 $\eta _{\mu \mu } = \eta _{\mu}^2$     & $\eta _{\mu Q_0}=\eta _{\mu}\eta_{Q_0}$ & $\eta_{\mu Q_1}=\eta _{\mu}\eta_{Q_1}$ & $\eta_{\mu Q_2}=\eta _{\mu}\eta_{Q_2}$ &  $\ell$  &  $\bar{\rho }_0\,, \mu\text{m}$  \\
 \hline
 $3.8\times 10^{-14}$ & $5.2\times 10^{-13}$ & $3.29\times 10^{-13}$ & $8.23\times 10^{-10}$ &  $1000$ &  $0.3$ 
   \\
 \hline
 $\eta _{Q_{00}}=\eta_{Q_0}\eta_{Q_0}$ & $\eta _{Q_{01}}=\eta _{Q_0}\eta_{Q_1}$ & $\eta_{Q_{02}}=\eta _{Q_0}\eta _{Q_2}$ & $\eta_{Q_{11}}=\eta_{{Q_1}}\eta_{Q_1}$ & $\eta_{Q_{12}}=\eta_{{Q_1}}\eta_{{Q_2}}$ & $\eta_{Q_{22}}=\eta_{Q_2}\eta_{{Q_2}}$ \\
 \hline
 $7.11\times 10^{-12}$ & $4.5\times 10^{-12}$ & $1.13\times 10^{-8}$ & $2.85\times 10^{-12}$ & $7.13\times 10^{-9}$ & $1.78\times 10^{-5}$ \\
\end{tabular}
        \caption{Dimensionless parameters of the model which correspond to the Figure \ref{fig:spectral-curve-1mm}. The dynamical non-paraxial contribution $\eta_{\text{Q2}}$ is the biggest one.}
\label{tab:parameters-1mm}
\end{table}

 Studies of the radiation from classical beams show that horizontal and vertical beam spreading lead to some modifications of the spectral line \cite{haeberle1997smith}. The horizontal spreading of the beam shifts the spectral line towards lower frequencies while the vertical spreading results in the opposite shift. A combination of both spreading types results in a broadening of the spectral line. Here we show that quantum coherence of the wave packet may lead to \textit{a different behavior}. Namely, despite the vertical-horizontal spreading of the wave-packet, the resulting spectral line \textit{does not demonstrate a broadening} until the quadrupole-quadrupole corrections come into play, which is the case for long gratings with $N \gg N_{\text{max}}$ only. 
 
 Such a stabilization of the line width can be explained using \eqref{expr-dWeQ1-by-FF}, \eqref{expr-dWeQ2-by-FF} and properties of the function $F(\omega)$.
 First of all, instead of a full width at half maximum (FWHM) one can consider a full width between zeros of the spectral curves. Zeros of the interference contributions are defined by zeros of the kernel $F$ itself\footnote{In a vicinity of a zero $\omega_0$, $F(\omega_0)=0$, the kernel $F$ can be factorized as $F(\omega)=(\omega-\omega_0)F_\text{res}(\omega)$. $(\omega-\omega_0)$ is a real function therefore in \eqref{expr-dWeQ1-by-FF}, \eqref{expr-dWeQ2-by-FF} the same factorization can be applied to the radiation intensities}.
Therefore, all contributions \eqref{expr-dWee-by-FF}-\eqref{expr-dWmumu-by-FF} have the same full width between zeros of the spectral curves (in Fig. \ref{fig:spectral-curve-1mm} an example of these coinciding zeros near the spectral maximum is presented). This strongly restricts the possible broadening of the spectral line and at large $N$ all contributions \eqref{expr-dWee-by-FF}-\eqref{expr-dWmumu-by-FF} tend to have the same width. 

\begin{figure}[h!]
 \centering
 \includegraphics[width=.75\linewidth]{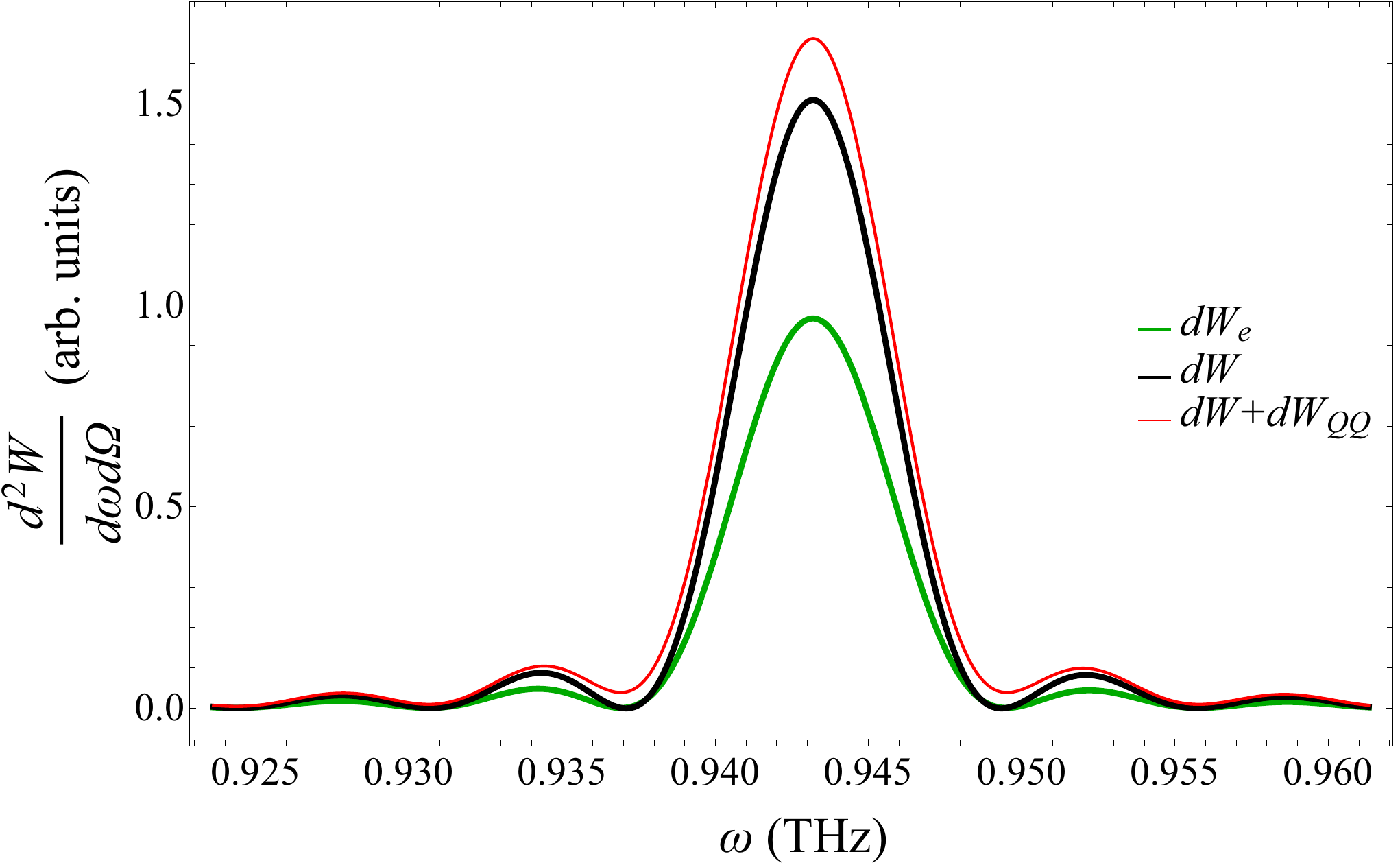}
 \caption{Broadening of the spectral line due to the quadrupole-quadrupole corrections to the S-P radiation. 
 The grating period $d=1\,$ mm, $\beta=0.5$, $a=d/2$, $\bar{\rho }_0=300 \text{nm}$, $\ell=1000$, $\Phi=\Theta=\frac{\pi}{2}$. Number of strips $N=150$. }
 \label{fig:QQ-domination-1mm}
\end{figure}

The quadrupole-quadrupole corrections $dW_{Q_1Q_2}$ and $dW_{Q_2Q_2}$ contain only derivatives of the kernel $F$. As a result, zeros of their spectral curves (in a vicinity of the maximum) disappear and the corresponding lines demonstrate a broadening (see figure \ref{fig:QQ-domination-1mm}). Importantly, if one takes into account these contributions then the next corrections of the same order should also be taken into account, such as interference of the charge with the octupole magnetic moment, with 16pole electric moment and so forth (see \eqref{def:radiation-intensity-interference-complete}). However, the octupole magnetic moment contribution has the same symmetry as the magnetic momentum contribution, thus  vanishing in the zenith direction. Regarding $dW_{e16p}$ contribution, because it is the interference term 
the zeros of the kernel $F$ will also prevent a broadening of the corresponding spectral line. As a result, the next corrections that may lead to a broadening are only the quadrupole-quadrupole ones. Therefore figure \ref{fig:QQ-domination-1mm} and table \ref{tab:width} contain all necessary terms.

Numerical studies of the spectral lines reveal not only an absence of the broadening, but even \textit{a slight narrowing} of the lines due to the charge-quadrupole interference. The FWHM for various grating lengths are presented in Table \ref{tab:width} where a narrowing of the line (Charge+LO corrections column) can be seen. When $N>100$ the quadrupole-quadrupole contributions become important (Charge+LO+NLO corrections column of Table \ref{tab:width}) and when $N>150$ broadening due to the horizontal-vertical spreading surpasses the narrowing. Thus, for $N<100$ we can safely compute $dW=dW_e+dW_{eQ_2}$ for parameters from Table \ref{tab:parameters-1mm}.

\begin{table}[]
    \centering
    \begin{tabular}{c|c|c|c}
 \text{} & \text{Charge} & \text{Charge+LO corrections} & \text{Charge+LO+NLO corrections} \\
 \hline
 \hline
 \text{$\Delta \omega $, THz, N=25} & 
 0.033408 & 0.033381 & 0.033383 \\
 \hline
 \text{narrowing, $\%$, N=25} & \text{} & $-0.081$ & $-0.077$ \\
 \hline
 \text{$\Delta \omega $, THz, N=50} & 0.016710 & 0.016658 & 0.016668 \\
 \hline
 \text{narrowing, $\%$, N=50} & \text{} & -0.31 & -0.25 \\
 \hline
 \text{$\Delta \omega $, THz, N=100} & 0.0083557 & 0.0082679 & 0.0083364 \\
 \hline
 \text{narrowing, $\%$, N=100} & \text{} &-1.05 & -0.23 \\
 \hline
 \text{$\Delta \omega $, THz, N=150} & 0.0055705 & 0.0054660 & 0.0056454 \\
 \hline
 \text{narrowing, $\%$, N=150} & \text{} & -1.8  & +1.3 (\text{broadening}) \\
 \end{tabular}
    \caption{Comparison of the FWHM for the charge radiation, with the interference terms included 
    (line narrowing) and with the quadratic terms included (line broadening). The grating period $d=1\,$ mm, $\beta=0.5$, $\bar{\rho}_0=0.3 \mu \text{m}$, $\ell=1000$, $h=0.39$ mm, $\Phi=\frac{\pi}{2}$,  $\Theta=\frac{\pi}{2}$}
    \label{tab:width}
\end{table}
A physical reason for the line narrowing is also spreading of the wave packet. Indeed, the natural width of the spectral line $\Delta \omega$ is related to the time scale of the radiation process $\Delta t$ by a following uncertainty relation:
\begin{equation}
    \Delta \omega \sim \frac{1}{\Delta t}\,. 
\end{equation}
Due to the packet spreading, its interaction with the strips lasts longer, especially at the end of the long grating with $dN_{\text{max}} \gg z_R$, and so $\Delta t(z)$ grows. In other words, the spreading slightly increases the radiation formation length.

\subsection{Angular distributions at the Smith-Purcell wavelength \label{sec:angular-distr-SPR}}

Let us denote 
$$
\omega_g = \frac{2\pi}{\lambda_g} =\frac{2\pi\, g}{d\left(\beta^{-1} - \cos \Theta\right)},\ g=1,2,3,...
$$
then $|F|^2$ contains a Fej\'{e}r kernel
\begin{eqnarray}\label{Fejer-kernel}
F_{N}(\omega)=\frac{\sin^2\left(\frac{N d\Theta_1}{2}\right)}{N \sin^2\left(\frac{d\Theta_1}{2}\right)} \xrightarrow[N\to\infty]{} 2\pi\sum_g \delta\left(d\omega(1/\beta-\cos\Theta)-d\omega_g(1/\beta-\cos\Theta)\right)=\cr
=\sum_g \frac{2\pi g}{g d(\beta^{-1}-\cos\Theta)} \delta\left(\omega-\omega_g\right)=\sum_g \frac{\omega_g}{g}\delta\left(\omega-\omega_g\right)
\end{eqnarray}
which can be used to integrate over frequencies in a vicinity of the resonant one. For the charge, the charge-dipole and the charge-$Q_0$ contributions, the Fej\'{e}r kernel can be substituted by a delta function when $N$ is large. For a grating of finite length with $N$ strips the spectral line has a width proportional to $1/N$. The angular distributions of the charge radiation and of the charge-dipole radiation for the main diffraction order $g=1$ read
\begin{eqnarray}\label{expr-dWee-SP}
&& \displaystyle
\frac{dW_{ee}}{ d\Omega}=N
\frac{d^2 \omega_1^3}{\pi^2}\sin^2\left(\frac{a\pi}{d}\right)
\exp\left(-\frac{2\omega_1 y}{\beta\gamma }\sqrt{1+\beta^2\gamma^2 \cos^2\Phi \sin^2\Theta}\right)
\cr
&& \displaystyle 
\times\frac{\cos^2\Theta+2\beta\gamma^2\cos^2\Phi\cos\Theta\sin^2\Theta+\sin^2\Phi\sin^2\Theta+\beta^2\gamma^4\cos^2\Phi\sin^4\Theta}{\beta^2\gamma^2(1+\beta^2\gamma^2 \cos^2\Phi \sin^2\Theta)},\\
\label{expr-dWemu-SP}
&& \displaystyle
\frac{dW_{e\mu}}{d\Omega}=
N \frac{\ell d^2 \omega_1^4}{m\pi^2}\,\frac{1}{\beta^2\gamma ^2}\,
\sin^2\left(\frac{\pi  a}{d}\right)
\exp \left(-\frac{2 \omega_1 y }{\beta\gamma}\sqrt{1+\beta^2 \gamma^2\cos^2\Phi \sin^2\Theta}\right)
\cr
&& \displaystyle 
\times\frac{\cos\Phi \sin\Theta \left(\beta\gamma^2 \sin^2\Theta+\cos\Theta\right)}{\sqrt{1+\beta^2 \gamma^2
   \cos^2\Phi \sin^2\Theta}}.
\end{eqnarray}
Both the intensities linearly increase with $N$.

Integration of the non-paraxial terms, $dW_{eQ_1}$ and $dW_{eQ_2}$, is more tricky. First we note that at large $N$ the spectral factors $F_{eQ_j}$ are concentrated near the Smith-Purcell frequency and have a width $\sim 1/N$. $F_{eQ_1}$ is approximately an odd function and $F_{eQ_2}$ is an even function of $\omega-\omega_1$ (see Fig. \ref{fig:spectral-curve-1mm}). Therefore $F_{eQ_1}$ produces a shift of the spectral maximum, whereas $F_{eQ_2}$ amplifies the intensity. At large $N$ these spectral factors are related with the  Fej\'{e}r kernel and its derivatives.

For instance, the charge-$Q_2$ intensity has the following factor at large $N$:
\begin{eqnarray}
& \displaystyle
F_{eQ_2}=2d^2N^3\Theta_1^2\sin^2\left[\frac{a \Theta_1}{2}\right]F_{N}(\omega) +{\rm O}(N)
\end{eqnarray} 
which is just proportional to the Fej\'{e}r kernel and can be substituted by a delta-function. 
As a result, the dynamically enhanced charge-quadrupole interference term $dW_{eQ_2}/ d\Omega$ reads 
\begin{eqnarray}\label{expr-dWeQ2}
& \displaystyle
\frac{dW_{eQ_2}}{ d\Omega}=\ell^2\left(\frac{\lambda_c}{\bar{\rho}_0}\right)^2\,\frac{1}{3\beta^4\gamma^4}\,\frac{f_2(N)}{\lambda_1^2}\,\frac{dW_{ee}}{ d\Omega}.
\end{eqnarray}
Expectedly, this dynamical contribution is suppressed in the relativistic case, $\gamma\gg 1$, 
when the spreading is marginal. Here
\begin{eqnarray}\label{expr-coherentQ2factor}
& \displaystyle
f_2(N) = 3\pi a d\, \cot \left( \frac{a\pi}{d} \right) + 3\pi^2\,a^2 + 3\pi\, a d (N - 1) + d^2\,(\pi^2(2 N^2-3 N + 1)-3) \approx \cr
& \displaystyle \approx 2\pi^2 d^2 N^2\ \text{when}\ N \gg 1.
\end{eqnarray}
Note that 
\begin{eqnarray}\label{f2N}
& \displaystyle
\frac{f_2(N)}{\lambda_1^2} \approx \frac{ 2\pi^2 d^2 N^2}{\lambda_1^2} \sim 2\pi^2 N^2\ \text{at}\ N \gg 1,\cr 
& \displaystyle \text{as}\ \lambda_1 \sim d\ \text{everywhere except}\ \Theta \to 0,
\end{eqnarray}
which leads to a cubic growth for the quadrupole-charge contribution  $dW_{eQ_2}/ d\Omega$ with respect to the number of strips ($dW_{eQ_2}/ d\Omega\sim N^2 dW_{ee}/ d\Omega$,  $dW_{ee}/ d\Omega\sim N$). The ratio of this correction to the radiation of the charge also has a non-linear (quadratic) $N$-dependence 
$$
\frac{dW_{eQ_2}}{ d\Omega}/\frac{dW_{ee}}{ d\Omega} \sim \ell^2\, N^2\,\left(\frac{\lambda_c}{\bar{\rho}_0}\right)^2,\ N \gg 1.
$$
Importantly, $1/\lambda_1^2(\Theta)$ is the only additional angle-dependent factor in $dW_{eQ_2}$ compared to $dW_{ee}$, so the dynamical contribution is increased for smaller wavelengths -- that is, for smaller emission angles, $\Theta \to 0$. Namely, at $\beta \approx 0.5$ we have
\begin{eqnarray}\label{dWeq2comp}
& \displaystyle
\frac{dW_{eQ_2} (\Theta = 0)}{dW_{eQ_2} (\Theta = \pi/2)} \approx 4.
\end{eqnarray}

A large $N$ asymptotic of the charge-$Q_1$ spectral factor reads
\begin{equation}
F_{eQ_1}=-N\omega\sin^2\left[\frac{a \Theta_1}{2}\right]F'_{N}(\omega) +{\rm O}(1)\,,
\end{equation}
where a derivative of the Fej\'{e}r kernel appears
$$
F'_{N}(\omega)=\frac{d \Theta_1 \sin \left(\frac{d N \Theta_1}{2}\right)\left[\cot\left(\frac{d \Theta_1}{2}\right)\sin \left(\frac{d N \Theta_1}{2}\right)-N\cos\left(\frac{d N \Theta_1}{2}\right) \right] }{N\omega \sin^2 \left(\frac{d\Theta_1}{2}\right)}
\xrightarrow[N\to\infty]{} \sum \frac{\omega_g}{g}\delta'\left(\omega-\omega_g\right).
$$
Integration of \eqref{expr-dWeQ12-factorized-structure} when $g=1$ can be done using a substitution of $F'_{N}(\omega)$ by a derivative of a delta-function 
$$
\frac{dW_{eQ_1}}{ d\Omega}=-N\omega_1\int 
\exp \left(\frac{-2 y\omega }{\beta\gamma}\sqrt{1+\beta^2 \gamma^2\cos^2\Phi \sin^2\Theta}\right)
P_{eQ_j}(k_x,y,\omega)\omega\sin^2\left[\frac{a \Theta_1}{2}\right]\delta'\left(\omega-\omega_g\right)d\omega
=
$$
$$
N\omega_1 \partial_\omega\left. \left[
\exp \left(\frac{-2 y\omega }{\beta\gamma}\sqrt{1+\beta^2 \gamma^2\cos^2\Phi \sin^2\Theta}\right)
P_{eQ_j}(k_x,y,\omega)\omega\sin^2\left(\frac{a \Theta_1}{2}\right)\right]\right|_{\omega\to \omega_1}\,.
$$

Using explicit expressions of the radiation intensities one can isolate the dimensionless parameters related with the quadrupole contribution. From Eq.\eqref{expr-dWeQ2}, one can find $\eta_{Q_2}$, while $\eta_{Q_1}$ and $\eta_{Q_2}$ can be extracted from the ratios $dW_{eQ_j}/dW_{ee}$:
\begin{align}
& \frac{dW_{eQ_0}}{dW_{ee}}=  \frac{\bar{\rho }_0^2}{h_\text{eff}^2} P_1(\beta,\Theta,\Phi)
+ \ell^2 \frac{\lambda_c^2}{\bar{\rho}_0^2} P_2\left(\beta,\frac{h}{d},\Theta,\Phi\right)\,,
\label{expr-proof-of-diml-par-etaQ0}
\\
& \frac{dW_{eQ_1}}{dW_{ee}}
=
 \ell^2 \frac{\lambda_c^2}{\bar{\rho}_0^2}
  P_3\left(\beta,\frac{h}{d},\Theta,\Phi\right)\,,
   \label{expr-proof-of-diml-par-etaQ0}
\end{align}
where $P_{1,2,3}$ are some smooth functions. One can identify three parameters \eqref{def:eQ0-interf-term-parameter}, \eqref{def:eQ1-interf-term-parameter}. Note that $\frac{dW_{eQ_0}}{dW_{ee}}$ contains a linear combination of independent parameters $\eta_{Q_0}$ and $\eta_{Q_1}$, which we split for a convenience.     For ultra-relativistic energies, all corrections from the quadrupole radiation are suppressed.

\begin{figure}[h!]
\centering
    \subfloat[$\ell=1000$, $N=50$]
    {\label{fig:azimuthal-d1-l1000}\includegraphics[width=.5\linewidth]{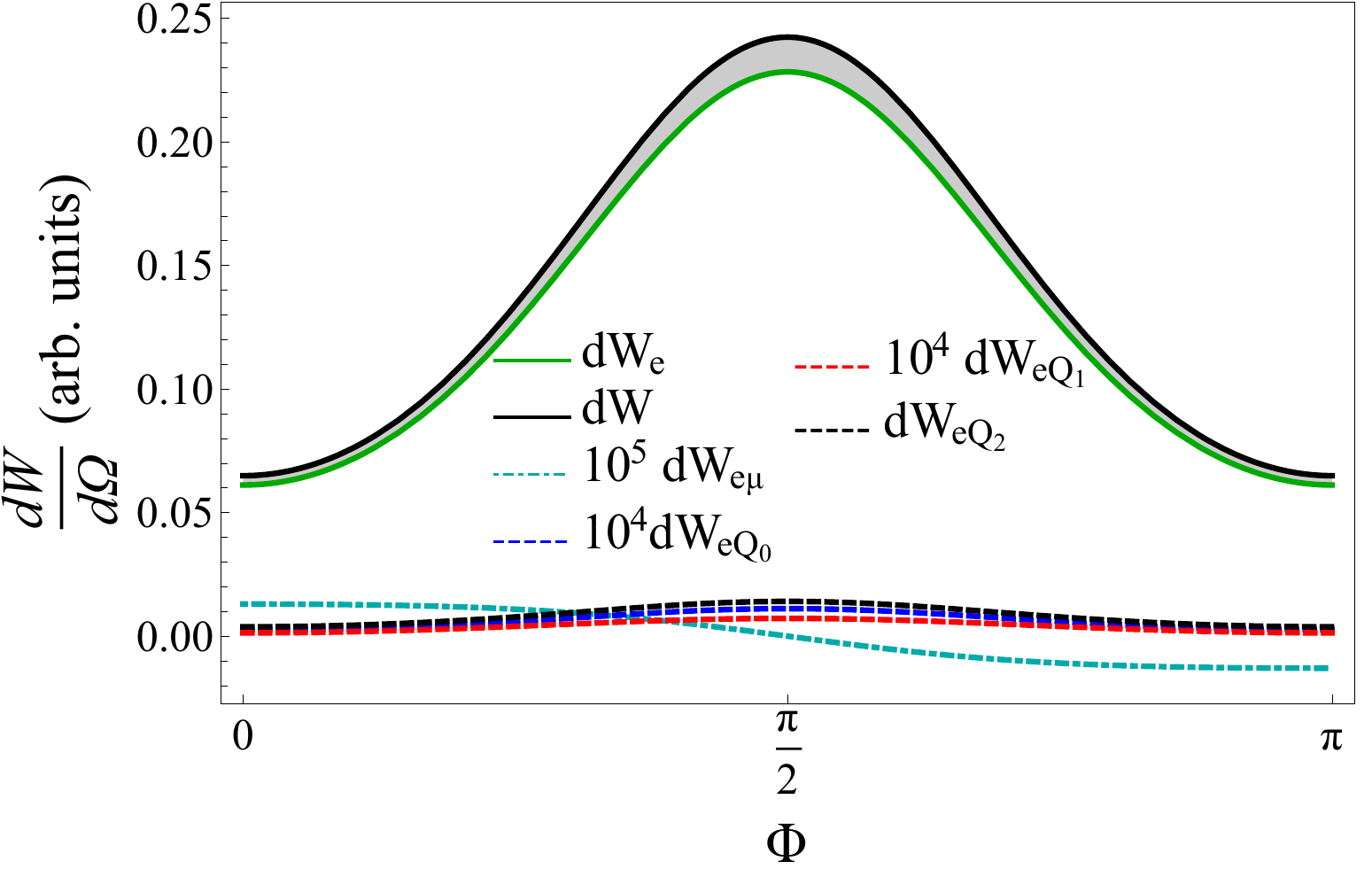}    }
    \\
    \subfloat[$\ell=100$, $N=500$]
    {\label{fig:azimuthal-d1-l100}\includegraphics[width=.5\linewidth]{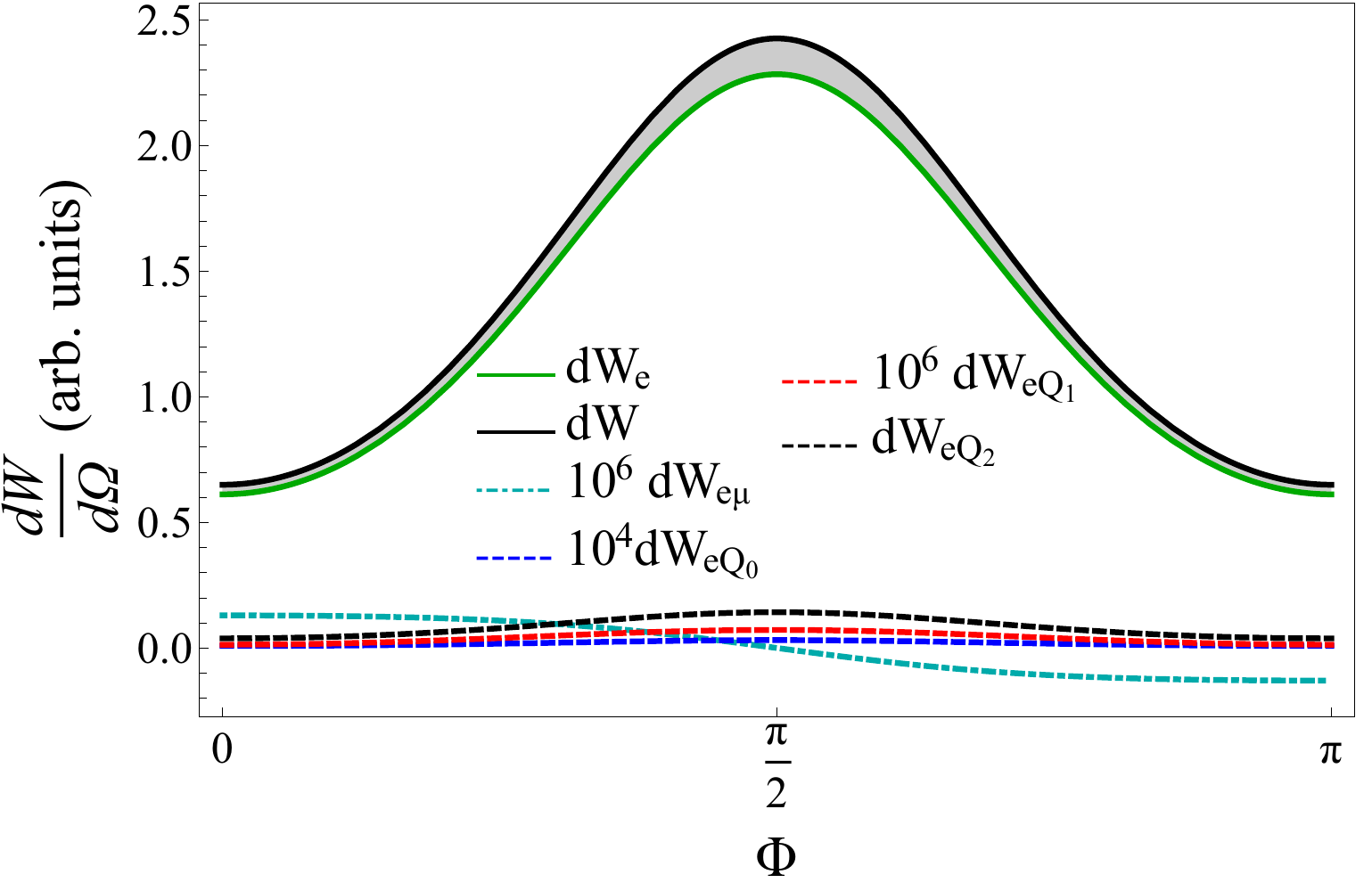}    }
    \caption{Azimuthal distribution of the radiation intensity (black solid line) and contributions from the charge (green solid line), the magnetic moment (cyan dashed line), and the electric quadrupole moment (black, red and blue dashed lines) with the following parameters: $\beta=0.5$, $\lambda=2\text{mm}$, $d=1 \text{mm}$, $a=d/2$. The number of strips $N$ is maximal in each subfigure for the given OAM,  the impact parameter $h=0.13$mm and $\bar{\rho}_0=0.3 \,\mu\text{m}$. }
    \label{azimutal-distribution}
\end{figure}

In Figure \ref{azimutal-distribution} we plot an azimuthal distribution of the radiation intensity at $\lambda=2\,\text{mm}$ 
and compare different contributions to the total radiation intensity. We fix the impact parameter and initial radius of the wave packet and consider two cases: $\ell=1000$, $N=50$ and $\ell=100$, $N=500$. In both cases the maximal number of strips  for a given impact parameter, OAM and the initial radius $\bar{\rho}_0$ is used. 
A larger grating length corresponds to the smaller angular momentum $\ell=100$.
A scaling invariance $\ell \to \epsilon \ell$, $N\to \epsilon^{-1}N$ of the quadrupole correction $dW_{eQ_2}/dW_{ee} \sim \eta_{Q_2}$ can be observed in Figure \ref{azimutal-distribution}. In other words, large OAM lead to a quick spreading and require short gratings, while small OAM result in a relatively slow spreading and allow one to use longer gratings. At the same time, other corrections -- in particular from the magnetic moment -- depend on $\ell$ only. Their observation requires the largest possible OAM ($\ell \sim 10^3$ and higher) and $\Phi \ne \pi/2$.  


Note that for $\beta=0.5$, the charge and the charge-quadrupole contributions have almost the same azimuthal dependence (which is defined mostly by the exponential factor). The charge-magnetic moment contribution yields the small azimuthal asymmetry. The analysis of Ref.\cite{ivanov2013detecting} seems reasonable for the case of the Smith-Purcell radiation too. In the case of THz radiation ($\lambda\sim 1\text{mm}$) this effect is almost unobservable (see figure \ref{azimutal-distribution} and Table \ref{tab:parameters-1mm}). An asymmetry of the order of 0.1 \% can be seen for infrared S-P radiation, $\lambda\sim 1\,\mu\text{m}$, which could in principle be measured. 



\subsection{Dynamical enhancement of the quadrupole contribution \label{subsec:coherent-effect}}

Equations \eqref{expr-dWeQ2}, \eqref{expr-coherentQ2factor} show that $dW_{eQ_2}$ contribution has a cubic growth with the number of strips compared to the linear growth of $dW_{ee}$. This is due to constructive interference of the quadrupole radiation from each strip, taken into account that the quadrupole moment is increased (quadratically) because of the spreading. Recall that the maximal grating length (interaction length) and the number of strips are limited by \eqref{N-max-expression} to guarantee that the mean wave packet radius $\bar{\rho}(t)$ stays smaller than the impact parameter $h$. 


A large impact parameter $h$ should be chosen to obtain a large $N_{\text{max}}$, therefore we consider $\bar{\rho}_0/h\ll 1$ and thus approximately $N_\text{max}^3\propto h^3$. At the same time, the radiation intensity decreases exponentially with the large impact parameters. Taking the maximal number of strips, the dependence of the charge-quadrupole interference term $dW_{eQ_2}$ on the impact parameter reads 
\begin{equation}
dW_{eQ_2}\left[\rho,\ell,h,N_\text{max}(\rho,\ell,h)\right]\propto \frac{\bar{\rho}_0}{\ell} \left(\frac{d }{ \lambda_c h_\text{eff}^5}\right) h^3\,
{\rm e}^{-\frac{2h}{h_\text{eff}}}\,.
\label{dWeQ2-optimal-rho-l-dependence}
\end{equation}
The maximum of this contribution defines the optimal impact parameter    
\begin{equation}\label{def:optimal-impact}
h_{\text{opt}}\approx \frac{3}{2}\, h_{\text{eff}} \sim h_{\text{eff}}.
\end{equation}
Note that $dW_{eQ_2}$ is proportional to $\bar{\rho}_0/\ell$ when $N$ takes its maximal value  \eqref{N-max-expression}. That is, a wide packet with a small OAM (recall the corresponding lower bound \eqref{lower-ell-bound}) can be chosen to simplify experimental studies of the S-P radiation from the vortex electrons.     

\begin{figure}[h!]
 \centering
 \includegraphics[width=.75\linewidth]{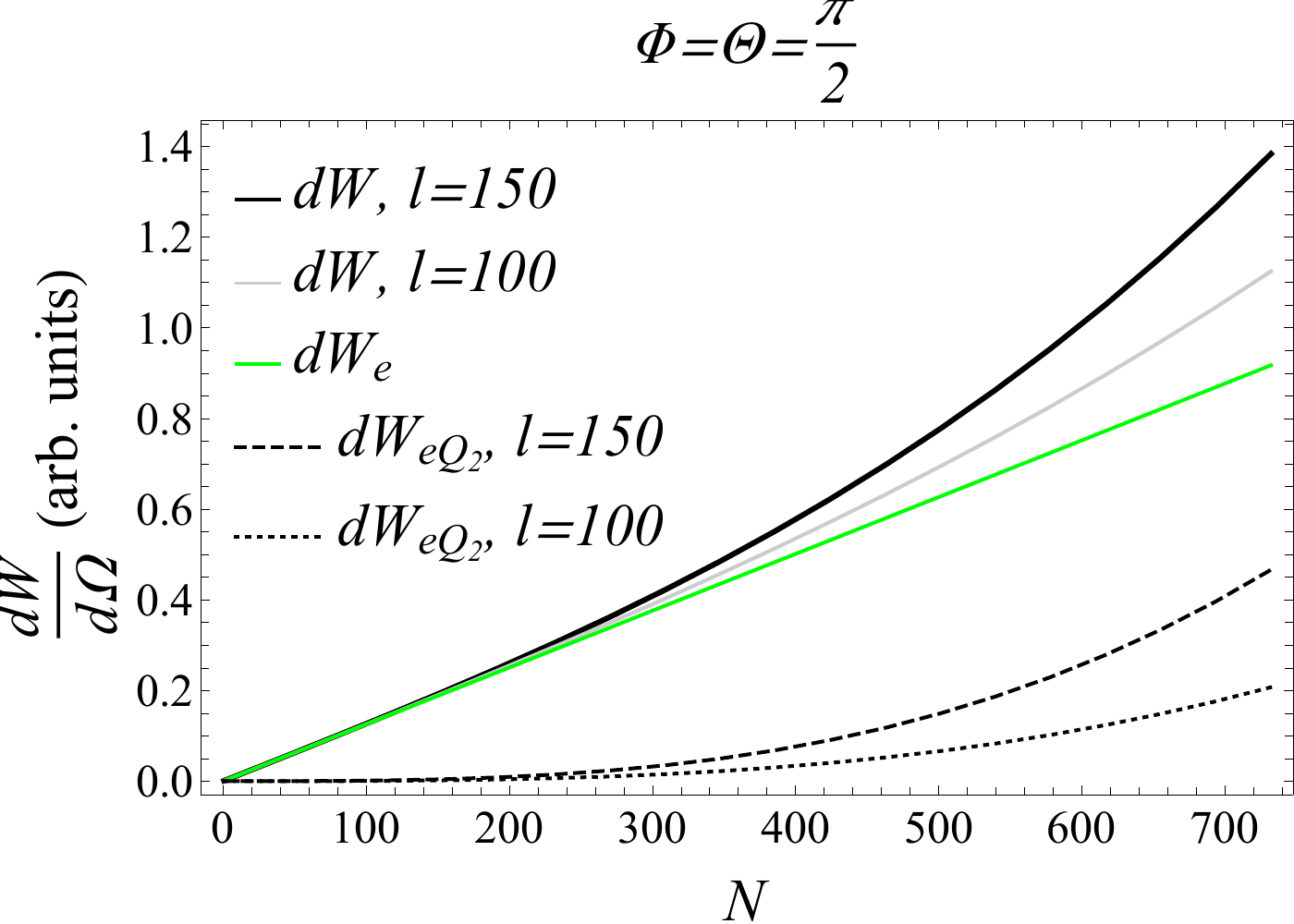}
 \caption{Radiation intensity at vertical plane for various number of strips $N<700$. The number of strips $N=700$ is maximal given that the impact parameter $h=3 h_\text{eff}/2=27\,\mu\text{m}$, velocity $\beta=0.5$, period $d=0.1\,$ mm and the initial mean radius $\bar{\rho}_0=300\,$nm.}
 \label{fig:dynamical-enchancement}
\end{figure}

In figure \ref{fig:dynamical-enchancement} the behavior of the radiation intensity is shown for the optimal value of $h$ and $N_{\text{max}}(h)=700$. Two cases of OAM, $\ell=150$ and $\ell=100$, were considered. The maximal number of strips was calculated for the larger OAM. The diffraction time is inversely proportional to $\ell$, therefore a grating optimal for a wave packet with $\ell=150$ can be used 
also in the case of $\ell<150$. In this case a factor $\ell^2/150^2$ reduces the charge-quadrupole radiation intensity $dW_{eQ_2}$.

The non-linear dependence of the S-P radiation on the number of strips or on the length of the grating due to the increasing quadrupole moment is clearly seen in figure \ref{fig:dynamical-enchancement}. If observed experimentally, such a non-linear dependence would serve as a hallmark for a non-paraxial regime of electromagnetic radiation, in which an electron packet emits photons as if its charge were smeared over all the coherence length, somewhat similar to a multi-particle beam but with a total charge $e$. Another possibility to detect this effect is to change OAM for the same diffraction grating and the same scattering geometry and to study corresponding modifications of the radiation intensity.

\begin{figure}[h!]
 \centering
 \includegraphics[width=.75\linewidth]{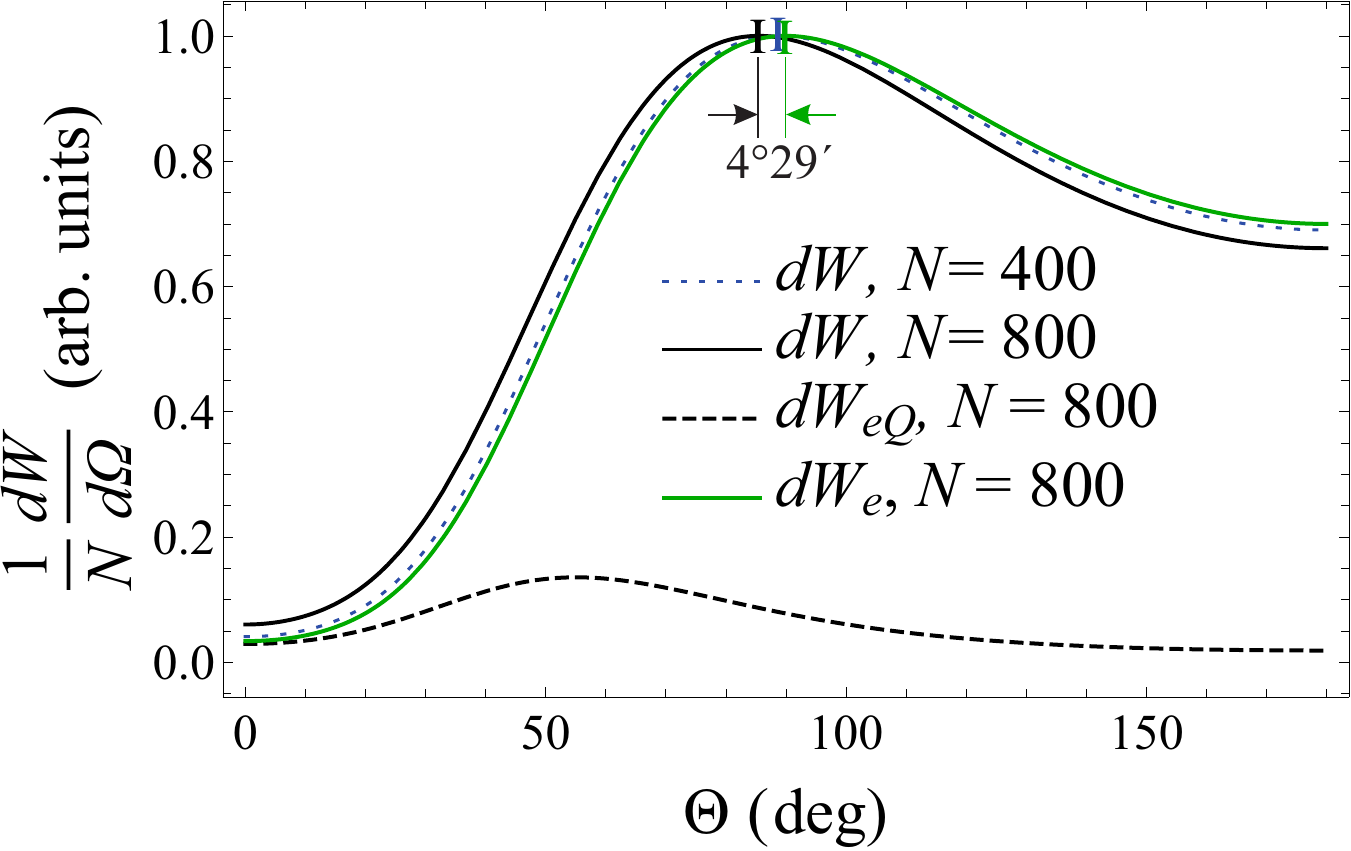}
 \caption{Polar dependence of radiation intensity and a shift of the maximum due to the contribution from $dW_{eQ_2}$. The number of strips $N=800$ is maximal given that the impact parameter $h=h_{eff}=22\,\mu\text{m}$, velocity $\beta=0.7$, period $d=0.1\,$ mm, OAM $\ell=10$ and the initial mean radius $\bar{\rho}_0=20\,$nm.}
 \label{fig:polar-dependence}
\end{figure}

An additional possibility to detect the charge-quadrupole contribution follows from a polar dependence in \eqref{expr-dWeQ2}. In most cases, the total radiation intensity is approximately $dW=dW_e+dW_{eQ_2}$ because other corrections are at least 2 orders of magnitude smaller than $dW_{eQ_2}$. In the case of large $N$ and $\Phi=\frac{\pi}{2}$, the maximum of the radiation intensity with respect to the polar angle $\Theta$ can be found by maximization of the following expression:
\begin{equation}
dW \propto  {\rm e}^{\frac{-2h}{h_\text{eff}}} \left(h_\text{eff}^{-3}+ \delta h_\text{eff}^{-5}\right) \,,\quad \delta=\frac{\eta_{Q_2}}{6}\frac{d^2}{\gamma^2\beta^2}\,.
\end{equation}
To begin with the polar angle of maximal intensity for the charge radiation $\Theta_e$ we just put $\delta=0$ which gives $h_\text{eff}(\Theta_e)=2h/3$ from a linear equation. A non-zero $\delta$ leads to the cubic equation. We use Cardano's formula and assume $\delta$ small to calculate the first order correction
\begin{equation}
h_\text{eff}(\Theta)=h_\text{eff}(\Theta_e)-\frac{\delta}{h}\,.
\end{equation}
Next, we write effective impact parameters explicitly in terms of the polar angle
$$
\frac{\beta\gamma d}{2\pi}(\cos(\Theta_e)-\cos(\Theta_e+\delta\Theta))=-\frac{\delta}{h}
$$
and for small shifts of the maximum, $\cos \delta\Theta \sim 1 $, $\sin \delta\Theta \sim  \delta\Theta $ 
we get an estimate
\begin{equation}\label{expr-polar-shift}
 \delta\Theta=-\eta_{Q_2}\frac{\pi d}{3h\gamma^3\beta^3 \sin(\Theta_e)}
\end{equation}
In figure \ref{fig:polar-dependence} we plot an example of this effect. The approximate shifts  $\delta\Theta(N=800)=-4.64^{\circ} $, $\delta\Theta(N=400)=-1.16^{\circ} $ from \eqref{expr-polar-shift} and numerical calculations $\delta\Theta(N=800)=-4.48^{\circ} $, $\delta\Theta(N=400)=-1.15^{\circ} $ are in a good agreement. 

\section{Conclusions} \label{sec:conclusions}
 We calculated the Smith-Purcell radiation generated by a vortex electron from an ideally conducting diffraction grating in the wavelength range from THz to optical range. The state of an electron is given by the Laguerre-Gaussian wave-packet with the orbital angular momentum. The corresponding intrinsic magnetic dipole moment and electric quadrupole moment were taken into account. The latter increases due to the “spreading” of the quantum wave-packet with time.

The radiation of the magnetic moment has an azimuthal asymmetry, as for transition radiation \cite{ivanov2013detecting}. The asymmetry effect increases with an increase in the orbital momentum of the wave packet, and for the currently achieved OAM values $\ell\sim 10^3$ it does not exceed one percent. 

In a contrast to the radiation of the magnetic moment, the azimuthal asymmetry is absent in the radiation of the electric quadrupole moment. However, as the electron moves near the grating, the spreading of the packet leads to an increase of the quadrupole moment, which also can be seen in the radiation. Although the quadrupole radiation is small as long as the multipole expansion stays legitimate (almost always in practice), it leads to an interesting effect: while the radiation intensity from a charge has a linear growth with the number of the grating periods, the quadrupole contribution leads to a faster cubic growth, which resembles coherence effect (superradiance) from a classical beam of many particles. However, in our problem this is a purely quantum effect of the spatial coherence of a vortex packet. For relativistic particles, spreading of the quadrupole moment can be neglected,  but for non-relativistic and even moderately relativistic electrons (with kinetic energies of ~ 100-300 keV), this effect can lead to both a change in the angular distribution and an increase in the total radiation loss. 

Thus we have shown that the effects of spatial coherence of the wave packets with intrinsic angular momentum can play a noticeable role for nonrelativistic energies of the radiating particles. Moreover, our calculations show that experimental observations of the quadrupole contribution to the S-P radiation can be done with a moderate value of OAM, $\ell\sim 100$. From Eq. \eqref{dWeQ2-optimal-rho-l-dependence} it follows that to study the dynamical effect which comes from the wave packet spreading it is better to work with small OAM, choose an optimal impact parameter \eqref{def:optimal-impact} and provide the maximal grating length \eqref{N-max-expression}. For instance, a $4^\circ$ shift of the polar angle of the maximum of Smith-Purcel radiation ($\lambda=0.14\,$mm) presented in figure \ref{fig:polar-dependence} can be achieved with $\ell=10$ and $\bar{\rho}_0=20\,$nm. Vortex electrons with such parameters can be generated using the method realized by J.  Verbeeck et. all  ( $|\ell|=3$, $\bar{\rho}_0~0.5-10\,$nm, see Fig. 2 in  \cite{verbeeck2011atomic}).
The resulting corrections can be detected just with a standard experimental setup as in \cite{remez2019observing} upgraded to work with vortex wave packets. Since the same shift should appear for LG wave packet with re-scaled parameters $\tilde\ell$, $\bar{\rho}_0=20 (\tilde\ell/\ell)\,$nm a possible experiment can choose between the focusing and vorticity.

\section*{Acknowledgements} \label{sec:acknowledgements}
 We are grateful to A.A. Tishchenko and P.O. Kazinski for fruitful discussions. This work is supported by the Russian Science Foundation (Project No. 17-72-20013).  
\section{Appendix} \label{sec:appendix}
   
\subsection{Electromagnetic fields of LG wave packet in the rest frame}  
 
Consider the vortex electron described by the LG packet \eqref{LGpsi} with $n=0$. Its electromagnetic fields represent a sum of those of the charge $e$, of the magnetic moment ${\bm \mu}$, and of the electric quadrupole moment $Q_{\alpha\beta}$, \eqref{Moments}. The fields in cylindrical coordinates in the rest frame were calculated in \cite{karlovets2019dynamical}. In our problem, we prefer to use the Cartesian coordinates: 
\begin{eqnarray}
&& 
E_x=\frac{x}{r^3}\left[1+\frac{1}{4}\left(\frac{3\bar{\rho}_0^2}{r^2}\left(1-5\frac{z^2}{r^2}\right)+\frac{l^2 \lambda_C^2}{\bar{\rho}_0^2}\left(\frac{3t^2}{r^2}\left(1-5\frac{z^2}{r^2}\right)+3\frac{z^2}{r^2}-1\right)  \right)\right],
\cr
&& 
E_y=\frac{y}{r^3}\left[1+\frac{1}{4}\left(\frac{3\bar{\rho}_0^2}{r^2}\left(1-5\frac{z^2}{r^2}\right)+\frac{l^2 \lambda_C^2}{\bar{\rho}_0^2}\left(\frac{3t^2}{r^2}\left(1-5\frac{z^2}{r^2}\right)+3\frac{z^2}{r^2}-1\right)  \right)\right],
\cr
&& 
E_z=\frac{z}{r^3}\left[1+\frac{1}{4}\left(\frac{3\bar{\rho}_0^2}{r^2}\left(3-5\frac{z^2}{r^2}\right)+\frac{l^2 \lambda_C^2}{\bar{\rho}_0^2}\left(\frac{3t^2}{r^2}\left(3-5\frac{z^2}{r^2}\right)+3\frac{z^2}{r^2}-1\right)  \right)\right],
\cr
&&
H_x = \frac{z}{r^5} \left(3x \frac{l}{2m} - \frac{3l^2 \lambda_C^2}{2\bar{\rho}_0^2} ty\right),
\cr
&&
H_y = \frac{z}{r^5} \left(3y \frac{l}{2m} - \frac{3l^2 \lambda_C^2}{2\bar{\rho}_0^2} tx\right),
\cr
&&
H_z = \frac{l}{2 m}\left(3 \frac{z^2}{r^2} - 1\right)\frac{1}{r^3}
\label{F-cartesian-comp}
\end{eqnarray}

We now transform these fields to the laboratory frame in which the particle moves along the $z$ axis with a velocity $\langle u \rangle \equiv \beta$ according to the law
$$
\langle z\rangle = \beta t
$$

Applying Lorentz transformations we get electric fields in the laboratory frame 
\begin{eqnarray}
&& \displaystyle E^{(\text{lab})}_{x} = \gamma  (E_{x} + \beta H_{y}),\cr
&& \displaystyle E^{(\text{lab})}_{y} = \gamma (E_{y} - \beta H_{x})\,,\ E^{(\text{lab})}_z = E_z
\label{FLabtrans}
\end{eqnarray} 
Simultaneously, we need to transform the coordinates and the time as follows \footnote{Note that Ref. \cite{karlovets2019dynamical} treats the fields at a distant point only, which simplifies Lorentz transformations of angular variables. Here we use the general formulas.}:
\begin{align}
 {\bm \rho} = \{x,y\} = \text{inv},\ z\rightarrow  \gamma(z - \beta t)=:R_z,\quad t \rightarrow \gamma(t-\beta z)=:T_z ,\cr
 r^2  \rightarrow  \left(\rho^2 +  \gamma^2(z - \beta t)^2\right)\,,\quad \gamma=(1-\beta^2)^{-1/2}
\label{trans}
\end{align} 
We omit the magnetic fields, as to calculate the surface current below we need the electric field only.

\subsection{Electromagnetic fields of LG wave packet in the laboratory frame}

Let's introduce a vector ${\bm R} = \{{\bm \rho}, \gamma (z - \beta t)\}$  in the laboratory frame.
The components of the electric field in this frame read 
\begin{eqnarray}
&& \displaystyle E_x({\bm r}, t) = \gamma\frac{x}{R^3} \Bigg(1 + \frac{3}{4} \frac{\bar{\rho}_0^2}{R^2} \left(1 - 5\frac{R_z^2}{R^2}\right) + \frac{1}{4} \ell^2 \left(\frac{\lambda_c}{\bar{\rho}_0}\right)^2 \Big[3 \frac{T_z^2}{R^2}\left(1 - 5\frac{R_z^2}{R^2}\right) + \cr&& \displaystyle + 3 \frac{R_z^2}{R^2} - 6\beta \frac{R_z T_z}{R^2} - 1\Big]\Bigg) + \frac{\ell}{2m} 3\beta\gamma y \frac{R_z}{R^5},\cr
&& \displaystyle E_y({\bm r}, t) = \gamma\frac{y}{R^3} \Bigg(1 + \frac{3}{4} \frac{\bar{\rho}_0^2}{R^2} \left(1 - 5\frac{R_z^2}{R^2}\right) + \frac{1}{4} \ell^2 \left(\frac{\lambda_c}{\bar{\rho}_0}\right)^2 \Big[3 \frac{T_z^2}{R^2}\left(1 - 5\frac{R_z^2}{R^2}\right) + \cr&& \displaystyle + 3 \frac{R_z^2}{R^2} - 6\beta  \frac{R_z T_z}{R^2} - 1\Big]\Bigg) - \frac{\ell}{2m} 3\beta\gamma x \frac{R_z}{R^5},\cr
&& \displaystyle E_z({\bm r}, t) = \frac{R_z}{R^3} \Bigg(1 + \frac{3}{4} \frac{\bar{\rho}_0^2}{R^2} \left(3 - 5\frac{R_z^2}{R^2}\right) + \frac{1}{4} \ell^2 \left(\frac{\lambda_c}{\bar{\rho}_0}\right)^2 \Big[3 \frac{T_z^2}{R^2}\left(3 - 5\frac{R_z^2}{R^2}\right) + \cr&& \displaystyle + 3 \frac{R_z^2}{R^2} - 1\Big]\Bigg).
\label{Elab}
\end{eqnarray}



\end{document}